\documentclass[a4paper,fleqn]{cas-sc}

\usepackage[numbers,sort&compress]{natbib}
\usepackage{amsmath,amssymb}
\usepackage{booktabs}
\usepackage{longtable}

\usepackage{makecell}
\usepackage{caption}
\usepackage{multirow}
\usepackage{xcolor}
\usepackage{tikz}
\usetikzlibrary{trees,arrows.meta,positioning,shapes.geometric,shapes.symbols,shapes.misc,fit,calc,backgrounds,decorations.pathreplacing,patterns}
\usepackage{pgfplots}
\usepackage[section]{placeins}
\pgfplotsset{compat=1.17}
\usepackage{array}

\newcolumntype{P}[1]{>{\raggedright\arraybackslash}p{#1}}
\newcolumntype{Q}[1]{>{\centering\arraybackslash}p{#1}}

\newcommand{\yes}{\ensuremath{\checkmark}}
\newcommand{\partyes}{\ensuremath{\odot}}
\newcommand{\nope}{}

\newcommand{\cmark}{\ensuremath{\checkmark}}
\newcommand{\xmark}{\ensuremath{\times}}

\begin{document}
\let\WriteBookmarks\relax
\def\floatpagepagefraction{1}
\def\textpagefraction{.001}

\shorttitle{Adversarial Diffusion Across Modalities}
\shortauthors{A. Alotaibi and M. Ahmed}

\title[mode=title]{Adversarial Diffusion Across Modalities: A Fusion Survey of Attacks, Defenses, and Evaluation for Text, Vision, and Vision-Language Models}

\author[1,2]{Abrar Alotaibi}[orcid=0000-0003-1168-8050]
\ead{amotaibi@iau.edu.sa}
\cormark[1]
\credit{Conceptualization, Methodology, Investigation, Data Curation, Writing -- Original Draft, Visualization}

\author[1,3]{Moataz Ahmed}[orcid=0000-0003-0042-8819]
\ead{moataz@kfupm.edu.sa}
\credit{Writing -- Review \& Editing, Supervision, Project administration, Funding acquisition}

\address[1]{Information and Computer Science Department, King Fahd University of Petroleum \& Minerals, Dhahran, 31261, Saudi Arabia}
\address[2]{College of Computer Science and Information Technology, Imam Abdulrahman Bin Faisal University, Dammam, 31441, Saudi Arabia}
\address[3]{SDAIA-KFUPM Joint Research Center for Artificial Intelligence, King Fahd University of Petroleum \& Minerals, Dhahran, 31261, Saudi Arabia}

\cortext[1]{Corresponding author}


\begin{abstract}
Adversarial evaluation of AI systems has matured along four largely disconnected research tracks: diffusion-based attacks on text and large language models (LLMs), diffusion-based attacks on image classifiers, jailbreak pipelines against vision-language models, and diffusion-based input purification defenses. Each track has developed its own vocabulary, threat models, and benchmarks, with denoising diffusion models emerging as a shared generative mechanism whose recipes are now being actively ported between communities. This survey performs an information-fusion exercise at the meta-research level: we integrate these four tracks into a single conceptual framework with a unified taxonomy, evaluation criteria, and research agenda, with primary focus on the LLM-side slice. We catalog fifty published papers across four scope areas (text/LLM, image classifier, vision-language model, defense), plus four diffusion-LLM-as-victim entries and ten non-diffusion baselines that any new diffusion-based attack must be compared against. We propose a six-class taxonomy of diffusion roles in adversarial pipelines, augmented by a threat-model axis that records attacker knowledge, query budget, and target accessibility, and we apply a five-dimension evaluation framework (attack success rate, transferability, query budget, perplexity, defense-evasion) uniformly across modalities. The review adopts a dual attacker-defender perspective: alongside the attack catalog we cover four diffusion-based defenses that constitute the natural evaluation backdrop for any new attack. We provide a critical analysis that identifies five recurring weaknesses of the current LLM-side literature, and we close with a research agenda of open questions and concrete experimental designs. The companion catalog and the underlying spreadsheet are released alongside the paper. We are explicit that this is a narrative review with quality assessment, not a PRISMA-compliant systematic review, and we discuss the implications of this choice for replication.
\end{abstract}


\maketitle

\section{Introduction}
\label{sec:intro}

The deployment of large language models (LLMs) in healthcare, finance, education, and national-security adjacent applications has made the assessment of their safety properties a first-order concern \citep{achiam2023gpt4,touvron2023llama2,jabbar2025red}. Red teaming, the structured practice of probing a system with adversarial inputs to uncover vulnerabilities, has emerged as the dominant paradigm for this assessment \citep{perez2022red,ganguli2022red}. Existing red teaming methods for LLMs span a range of techniques, including discrete-token optimization of adversarial suffixes \citep{zou2023universal}, genetic-algorithm refinement of fluent jailbreaks \citep{liu2024autodan}, attacker-LLM iterative rewriting in black-box settings \citep{chao2023jailbreaking}, and autoregressive amortized attack generators \citep{paulus2024advprompter}. A complementary technological track has reshaped generative modeling over the past five years. Denoising diffusion models, which formulate generation as iterative noise removal, have become the state of the art for image synthesis \citep{ho2020denoising,song2021scorebased,rombach2022high}, and they are now being extended to discrete and text modalities through structured discrete-diffusion processes \citep{austin2021structured,lou2024discrete,shi2024simplified} and large-scale masked diffusion language models \citep{nie2025llada}.

The intersection of these two trajectories, diffusion models used as adversarial generators against LLMs, is much less developed than the broader LLM red teaming literature. Recent surveys of LLM red teaming \citep{jabbar2025red,lin2025red,yi2024jailbreak,cui2024risk} catalog hundreds of attacks, but diffusion-based attacks on text receive at most a sentence of mention. Conversely, surveys of diffusion models concentrate on image generation and rarely discuss adversarial usage. The result is that researchers entering this space have no consolidated reference describing what has been tried, what has worked, what has failed, and where the gaps lie.

This review aims to fill that gap. We focus on the use of diffusion models as the generative mechanism in adversarial pipelines that target machine-learning systems, with a particular emphasis on the LLM-side literature. We additionally cover the much larger image-domain literature because it provides the methodological vocabulary that text-side authors are now porting, and we cover diffusion-based defenses for text models because no diffusion-attack paper can be evaluated without them. The central observation is that the LLM-side literature is small but methodologically diverse: only four published papers operate in this regime as of writing, yet they cover four distinct attack philosophies and are appearing at an accelerating pace. The image-side literature, by contrast, is mature and contains specific recipes (DDIM-inversion plus latent perturbation; classifier-score guidance; segment-wise back-propagation through purifiers) that can be transplanted to the text setting.

The motivation for this work is threefold. First, the practical motivation is that as LLMs are deployed in higher-stakes settings, defenders need to understand the full landscape of attack mechanisms, including those that have not yet matured but plausibly will. Second, the methodological motivation is that diffusion-based generators offer specific properties (bidirectional conditioning, principled sampling diversity, joint modeling of input and output) that the dominant autoregressive attackers do not, and these properties suggest several attack designs that have not been published. Third, the disciplinary motivation is that the literature in this space sits across at least four communities (NLP safety, adversarial robustness in vision, multimodal LLM safety, generative modeling), and a unified treatment helps reduce duplicated effort across these communities. In doing so, the present review is itself an exercise in information fusion: we integrate four previously-disconnected literatures, each with its own vocabulary, threat models, and benchmarks, into a single conceptual framework. This cross-stream synthesis is a natural fit for the kind of multi-source, multi-process integration that the field of information fusion has long been concerned with.

The contributions of this review are as follows.
\begin{itemize}
\item \textbf{Propose a taxonomy of diffusion roles in adversarial pipelines.} We introduce a six-class taxonomy of how diffusion is used in an attack (trained generator; frozen with latent perturbation; frozen with score or classifier guidance; off-the-shelf inference; pipeline-only renderer; victim diffusion model), augmented by a threat-model axis (attacker knowledge, query access, target accessibility). The taxonomy distinguishes papers that exploit diffusion's gradients from those that use it as a black-box rendering step.

\item \textbf{Develop a unified five-dimension evaluation framework.} We introduce a five-criterion evaluation framework (attack success rate, transferability, query budget, perplexity, defense-evasion) and apply it uniformly across text, image-classifier, and vision-language-model attacks (Section~\ref{sec:quantitative}). The framework pulls together reported figures from cataloged methods on shared benchmarks and exposes the closed-frontier-model gap that diffusion-based attacks have not yet closed.

\item \textbf{Conduct a cross-modality fusion of fifty papers across four streams.} We catalog fifty published papers spanning four previously-disconnected research streams (text/LLM red teaming, image-classifier adversarial machine learning, vision-language-model jailbreaking, and diffusion-based defenses), alongside four diffusion-LLM-as-victim entries and ten non-diffusion baselines. Each paper is recorded with its diffusion role, formulation, training method, datasets, metrics, target models, threat model, and code availability. The catalog and its underlying spreadsheet are released alongside the manuscript and are the source of all cross-cutting tables in this paper.

\item \textbf{Investigate the dual attacker-defender landscape.} Alongside the attack catalog, we cover four diffusion-based defenses (three text-classifier defenses and one multimodal-LLM defense) that constitute the natural evaluation backdrop for any new diffusion-based attack. We identify five recurring weaknesses of the current text-side attack literature and substantiate the analysis with the quantitative comparison built from our evaluation framework.

\item \textbf{Identify open research questions with concrete experimental designs.} We close with a set of open research questions grouped by methodological scope, framed as opportunities for the community rather than as a punch list of specific contributions. We acknowledge that some of these questions may already be under unpublished investigation at the time of reading.
\end{itemize}

The rest of this article is organized as follows.
Section~\ref{sec:method} describes the review method.
Section~\ref{sec:background} provides terminological background
and compares this review to existing surveys.
Section~\ref{sec:taxonomy} presents the taxonomy of diffusion-based
attacks and discusses each branch in detail.
Section~\ref{sec:defenses} covers diffusion-based defenses for text
models. Section~\ref{sec:quantitative} synthesizes reported empirical
results across the cataloged methods using our five-dimension
evaluation framework. Section~\ref{sec:evaluation}
summarizes evaluation methods and benchmarks.
Section~\ref{sec:discussion} returns to the research questions and
answers each one drawing on the catalog.
Section~\ref{sec:threats} states the limitations and threats to
validity of the review. Section~\ref{sec:implications} addresses
implications and forward research directions.
Section~\ref{sec:conclusion} concludes.

\section{Review Method}
\label{sec:method}

Our review method follows the structured process used in the broader red teaming literature \citep{jabbar2025red}, comprising search-strategy design, study selection, quality assessment, and data synthesis. We are explicit at the outset that this is a narrative review with quality assessment rather than a PRISMA-compliant systematic review \citep{page2021prisma}: the authors performed the complete screening pipeline (identification, title-and-abstract screening, full-text eligibility review, quality assessment, and final inclusion decisions) by hand over the course of the review, and we additionally report a supplementary post-hoc agreement check between two independent LLM agents on the inclusion rule (Section~\ref{sec:selection}). We did not pre-register the protocol or conduct independent dual-author blinded screening of the full pool with formal inter-rater agreement; the choice reflects the small size of the directly-on-target corpus (the LLM-side cluster contains four papers) and the fast pace of the field, which together would make formal systematic-review machinery disproportionate to the available evidence. The complete screening log, including titles, screening decisions, and exclusion reasons for every record considered, is available in the companion repository for this review: \url{https://github.com/AbrarAlotaibi/diffusion-redteam-llm-survey}.

\subsection{Research Questions}
\label{sec:rqs}
The review is guided by the following research questions, which are deliberately narrower in scope than questions used by general red teaming surveys because the topic itself is narrower.
\begin{itemize}
\item \textbf{RQ1.} What is the current state of research on the use of diffusion models for adversarial attacks on machine-learning systems, with particular emphasis on LLMs?
\item \textbf{RQ2.} What taxonomy can be used to classify the role of diffusion models in adversarial pipelines?
\item \textbf{RQ3.} What formulations, training methods, and optimization strategies have been used to produce adversarial outputs through diffusion?
\item \textbf{RQ4.} What datasets, metrics, and target models are used in the evaluation of diffusion-based attacks?
\item \textbf{RQ5.} What diffusion-based defenses have been proposed for text models, and how do they relate to the attack literature?
\item \textbf{RQ6.} What gaps and unaddressed questions emerge from the current literature, and which of these gaps are most tractable for follow-on research?
\end{itemize}

\subsection{Search Strategy}
\label{sec:search-strategy}
We conducted searches across arXiv, Google Scholar, OpenReview, the ACL Anthology, the IEEE Xplore Digital Library, and the ACM Digital Library, supplemented by the proceedings of the major machine-learning venues (NeurIPS, ICML, ICLR), computer-vision venues (CVPR, ICCV, ECCV), natural-language-processing venues (ACL, EMNLP, NAACL), and computer-security venues (USENIX Security, IEEE S\&P, ACM CCS, NDSS). Queries combined the keywords ``diffusion model'' or ``score-based'' with the keywords ``adversarial,'' ``jailbreak,'' ``red teaming,'' ``attack,'' and ``adversarial example,'' optionally restricted by modality (``text'', ``vision-language model'', ``image classifier''). References and citation chains of identified papers were then traversed recursively to surface adjacent work that did not appear in the keyword search.

\subsection{Study Selection and Quality Assessment}
\label{sec:selection}
The initial search returned 154 candidate records after cross-database de-duplication. We performed title-and-abstract screening on the full pool, which reduced it to 94 records that plausibly used diffusion in an adversarial pipeline; we then conducted full-text eligibility review on those 94 records, which reduced the corpus to the 50 papers retained in the catalog of this review (comprising four diffusion-attack papers on text/LLMs, eighteen on image classifiers, ten on vision-language models, four on diffusion-LLM-as-victim, four diffusion-based defenses, and ten non-diffusion baselines). The remaining 44 records that passed title-and-abstract screening but were excluded at full-text review are documented with exclusion reasons in our screening log, released as part of the companion repository (Section~\ref{sec:data-synthesis}).

The principal exclusion rule was that diffusion be a constitutive component of an adversarial pipeline, either as the generator of the adversarial input, the victim of the attack, or an explicit defense built on a diffusion-style denoising process. Papers that mentioned diffusion only in passing, used diffusion only for benign purposes (e.g., image generation for non-adversarial dataset construction), or whose use of the term ``diffusion'' was metaphorical (e.g., information diffusion in social networks) were excluded. Quality assessment of retained papers used the checklist in Table~\ref{tab:quality}, which adapts the criteria of the prior LLM red teaming review~\citep{jabbar2025red} to the diffusion-attack setting. The full pipeline: search, identification, title-and-abstract screening, full-text eligibility review, quality assessment against the checklist, and final inclusion decisions, was performed by the authors, with the first author conducting the bulk of the screening work under the supervision of the second author. Disagreements at the full-text stage were resolved through discussion to consensus.

\begin{table}[!htbp]
\caption{Quality-assessment checklist used during paper screening.}
\label{tab:quality}
\small
\centering
\begin{tabular}{c P{13.4cm}}
\toprule
S.~No. & Quality assessment question \\
\midrule
Q1 & Is the diffusion role precisely defined (trained, frozen, score-guided, pipeline-only)? \\
Q2 & Is the formulation (continuous, discrete, latent, embedding-space) clearly stated? \\
Q3 & Are the evaluation datasets and metrics fully reported? \\
Q4 & Are baseline comparisons against non-diffusion adversarial attacks provided? \\
Q5 & Does the study contribute to academic understanding of diffusion as an adversarial mechanism? \\
Q6 & Is the venue or citation visibility consistent with substantive peer review? \\
Q7 & Is target-model coverage sufficient (open-source, closed-source, defended)? \\
Q8 & Is code or a reproducible artifact released? \\
\bottomrule
\end{tabular}
\end{table}

After the authors' screening pipeline was complete, we additionally ran a supplementary post-hoc agreement check using two independent large-language-model agents to estimate how unambiguously the inclusion rule can be applied by a careful but uninformed reader. The two agents re-screened all 154 candidate records using only each record's title and a short summary, with distinct prompting strategies designed so that disagreements would reflect genuine rule ambiguity rather than within-model consistency: Agent~A applied the rule as a strict literal checklist with default-to-exclude on ambiguity, while Agent~B applied the rule with reasoning-first interpretation and default-to-include on ambiguity within the diffusion-plus-adversarial-plus-LLM/CV/VLM intersection. Cohen's $\kappa$ between the two agents was $0.781$, indicating substantial agreement under the criteria of \citet{landis1977measurement}; observed agreement was $89.6\%$ on a chance-corrected baseline of $52.5\%$. The two agents agreed with the authors' final catalog on all 50 inclusions and on 88 of 104 exclusions; the 16 boundary cases on which the two agents disagreed (late-find adjacent work such as RedDiffuser and VERA-V, defense variants such as DiffCAP, face-attack variants, and patch-defense work) represent the legitimate degrees of freedom in the catalog scope. We emphasize that this LLM-agent check is a supplementary transparency probe on the precision of the inclusion rule and is not equivalent to formal inter-rater agreement between independent human reviewers, which we did not conduct; the search, screening, quality assessment, and the final fifty-paper catalog are the authors' work. The LLM-agent prompts and re-screening decisions are released in the companion repository alongside the screening log.

\subsection{Data Synthesis}
\label{sec:data-synthesis}
For each retained paper we extracted the following information: title, authors, venue, year, arXiv identifier or DOI, scope area, diffusion role, diffusion type and space (continuous, discrete, latent, embedding, pixel, h-space, etc.), training or optimization method, dataset(s), metrics, target model(s), code availability, key contribution, and limitation. The extracted data are recorded in a relational catalog and released alongside this paper in a public GitHub repository (\url{https://github.com/AbrarAlotaibi/diffusion-redteam-llm-survey}), which contains the screening log, the per-paper metadata catalog, and the inclusion/exclusion decisions. All tables in this paper are generated from this catalog.

\subsection{Scope}
\label{sec:scope}
The scope of this review is the use of diffusion models as the generative mechanism in adversarial pipelines that target machine-learning systems, with the primary focus on LLMs. The image-domain literature (eighteen papers in our catalog) is included as methodological background because it predates and informs the LLM-side work. The vision-language-model literature (ten papers) is included because the LLMs in those settings are downstream consumers of diffusion-generated adversarial images. The dLLM-as-victim literature (four papers) is adjacent rather than in-scope, but is included because hybrid attacker designs may exploit those findings. Diffusion-based defenses (four papers: three text-classifier defenses and one multimodal-LLM defense) are included because they constitute the natural evaluation backdrop for any new diffusion-based attack. Non-diffusion baselines (ten papers) are included because every diffusion-based attack must be compared against them. Defenses for image models (DiffPure and follow-ups) are referenced where they bear on the attack literature but are not enumerated, because they fall outside the LLM-centered scope.

\section{Background}
\label{sec:background}

Before turning to the cataloged works themselves we provide two pieces of background. The first is the technical vocabulary of the diffusion-attack literature, which is non-trivial because the area draws from at least two distinct communities (denoising diffusion modeling on one hand, adversarial machine learning on the other) and the resulting terminology can be ambiguous. For example, the term ``guidance'' is used in the diffusion literature to mean classifier or classifier-free guidance applied to the score during sampling, whereas in the LLM safety literature it sometimes refers to instruction-following guidance or alignment guidance, with no direct connection to a score function. We collect the precise senses used in this paper in Section~\ref{sec:terminology}. Readers familiar with one community but not the other may wish to skim Table~\ref{tab:terms} before reading Section~\ref{sec:taxonomy}. The second piece of background is the set of prior review articles closest to the present work. Section~\ref{sec:related-reviews} compares this review to those surveys and identifies the specific gap that the present work addresses.

\subsection{Terminology}
\label{sec:terminology}
We provide brief definitions of the most frequently used terms in this review in Table~\ref{tab:terms}. The definitions emphasize how each term is used in the diffusion-attack literature specifically, which can differ from generic usage.

\begin{table}[!htbp]
\caption{Terminology used throughout this review.}
\label{tab:terms}
\centering
\small
\begin{tabular}{P{3.4cm} P{12cm}}
\toprule
Term & Definition \\
\midrule
Diffusion model & A generative model that learns to invert a forward noising process. The forward process gradually corrupts data with noise; the reverse process iteratively denoises samples from a noise distribution to the data distribution \citep{ho2020denoising,song2021scorebased}. \\
DDIM inversion & A deterministic inversion procedure for a pretrained denoising diffusion implicit model (DDIM) that maps a clean sample to its corresponding latent. DDIM inversion enables targeted editing or perturbation of a specific input by manipulating its inverted latent and re-denoising \citep{song2021denoising}. \\
Latent diffusion & A diffusion model whose forward and reverse processes operate in the latent space of a pretrained autoencoder rather than on raw pixels. Latent diffusion underlies Stable Diffusion and is the basis for most VLM-side adversarial work \citep{rombach2022high}. \\
Score-based / classifier guidance & Inference-time modification of the reverse process by adding a gradient term derived from a classifier. Used in the original guided-diffusion line~\citep{dhariwal2021diffusion,nichol2022glide} and adapted to inject adversarial signal at every denoising step \citep{dai2024advdiff,chen2023advdiffuser,huang2025scoreadv}. \\
Discrete diffusion & A diffusion process defined directly on discrete categorical data, such as token sequences. Examples include D3PM \citep{austin2021structured}, masked discrete diffusion (MDLM) \citep{shi2024simplified}, and score-entropy discrete diffusion (SEDD) \citep{lou2024discrete}. \\
Diffusion LLM (dLLM) & A large language model whose generation mechanism is non-autoregressive and based on iterative denoising or remasking, e.g., LLaDA \citep{nie2025llada}. Distinct from autoregressive LLMs in that prediction order is not strictly left-to-right. \\
Adversarial example & An input crafted to cause a target model to behave in an unintended way. In the LLM setting this is typically a prompt that elicits content the model is aligned to refuse. \\
Jailbreak & An attack that bypasses a model's safety training, causing it to comply with a request it would otherwise refuse \citep{wei2023jailbroken}. \\
Attack Success Rate (ASR) & The fraction of attempted attacks that succeed by some pre-specified judge, typically a keyword check, a learned classifier, or a strong-judge LLM such as Llama-Guard \citep{inan2023llamaguard}. \\
Transferability & The degree to which an attack crafted against one target model also succeeds against other target models. \\
\bottomrule
\end{tabular}
\end{table}

\subsection{Related Reviews}
\label{sec:related-reviews}
A number of recent surveys cover LLM safety and red teaming, but to the best of our knowledge none provides a focused treatment of diffusion-based adversarial generation. Table~\ref{tab:relatedreviews} situates the present review against eight recent works.

\citet{jabbar2025red} provide the most directly comparable prior survey, covering prompt-based attacks, data manipulation, model exploitation, information extraction, and model degradation. Their treatment of diffusion-based attack generation is limited to a single paragraph noting its existence, and the LLM-side text-diffusion attacks (DiffusionAttacker, DART) had not yet appeared at the time their survey was submitted. \citet{lin2025red} provide a complementary perspective from the generative-models side; their taxonomy categorizes attacks by exploited capability and includes some VLM jailbreak entries that we discuss here, but does not separate diffusion-based attacks as a methodological family. \citet{yi2024jailbreak} survey LLM jailbreak attacks and defenses but focus on prompt-engineering and gradient-based suffix attacks. \citet{cui2024risk} introduce a module-oriented risk taxonomy for LLM systems, and \citet{huang2024survey} survey safety and trustworthiness of LLMs through the lens of verification and validation, with diffusion-based attacks again receiving limited treatment. None of these reviews provides cross-modality coverage of diffusion-based attacks alongside LLM-specific text-diffusion work, which is the gap that the present review addresses.

\begin{table}[!htbp]
\caption{Comparison of this review with related surveys on LLM and adversarial-attack literature. \yes: covered, \partyes: partially covered, blank: not covered.}
\label{tab:relatedreviews}
\centering
\small
\begin{tabular}{P{3.0cm} c c c c c c c}
\toprule
Reference & \makecell{Diffusion as\\generator} & \makecell{Text /\\LLM} & \makecell{Image\\classifier} & \makecell{VLM /\\multimodal} & \makecell{dLLM as\\victim} & \makecell{Diffusion\\defense} & \makecell{Eval. \&\\benchmarks} \\
\midrule
\citeauthor{jabbar2025red}, \citeyear{jabbar2025red}             & \partyes  & \yes      & \nope     & \partyes  & \nope     & \nope     & \yes \\
\citeauthor{lin2025red}, \citeyear{lin2025red}                   & \partyes  & \yes      & \partyes  & \yes      & \nope     & \nope     & \partyes \\
\citeauthor{yi2024jailbreak}, \citeyear{yi2024jailbreak}         & \nope     & \yes      & \nope     & \partyes  & \nope     & \partyes  & \yes \\
\citeauthor{cui2024risk}, \citeyear{cui2024risk}                 & \nope     & \yes      & \nope     & \partyes  & \nope     & \nope     & \partyes \\
\citeauthor{huang2024survey}, \citeyear{huang2024survey}         & \nope     & \yes      & \partyes  & \partyes  & \nope     & \partyes  & \yes \\
\citeauthor{liu2024jailbreaking}, \citeyear{liu2024jailbreaking} & \nope     & \yes      & \nope     & \nope     & \nope     & \nope     & \partyes \\
\citeauthor{perez2022red}, \citeyear{perez2022red}               & \nope     & \yes      & \nope     & \nope     & \nope     & \nope     & \partyes \\
\citeauthor{ganguli2022red}, \citeyear{ganguli2022red}           & \nope     & \yes      & \nope     & \nope     & \nope     & \nope     & \yes \\
\midrule
\textbf{This review}                                             & \yes      & \yes      & \yes      & \yes      & \yes      & \yes      & \yes \\
\bottomrule
\end{tabular}
\end{table}


\section{Diffusion-Based Adversarial Attack Taxonomy}
\label{sec:taxonomy}

We organize the literature first by target modality (text/LLM, image
classifier, vision-language model) and within each modality by the role that the diffusion model plays in the adversarial pipeline. The fifty cataloged papers fall into five methodological families (A--E) plus a baseline set (F): four diffusion-attack papers on text/LLMs (Family A, Section~\ref{sec:text-attacks}); four diffusion-LLM-as-victim papers (Family B,
Section~\ref{sec:dllm-victim}, adjacent rather than primary); eighteen on image classifiers (Family D, Section~\ref{sec:image-attacks});
ten on vision-language models (Family E,
Section~\ref{sec:vlm-attacks}); ten non-diffusion baselines
(Family F, Section~\ref{sec:baselines}); and four diffusion-based defenses (Family C, Section~\ref{sec:defenses}).

The role-within-modality view makes the internal structure of each attack-modality visible (Figure~\ref{fig:taxonomy}). Family~A divides into a trained-generator regime (three papers) and an off-the-shelf inference regime (one paper). Family~D spans four distinct diffusion roles: latent perturbation,
score guidance, patch/physical, and purifier evasion. Family~E spreads across four further roles, including the only diffusion-as-victim paper in the catalog.

%
\begin{center}
\begin{minipage}{\linewidth}\centering
\begin{tikzpicture}[
  font=\sffamily\scriptsize,
  modality/.style={rectangle,rounded corners=2pt,draw=blue!50!black,fill=blue!12,
                   font=\sffamily\bfseries\small,align=center,inner sep=3pt,
                   minimum height=14mm,text width=18mm},
  family/.style={rectangle,rounded corners=2pt,draw=orange!60!black,fill=orange!10,
                 font=\sffamily\bfseries\scriptsize,align=center,inner sep=2.5pt,
                 minimum height=8mm,text width=22mm},
  papers/.style={rectangle,rounded corners=2pt,draw=gray!50,fill=gray!4,
                 font=\sffamily\scriptsize,align=center,inner sep=2.5pt,
                 minimum height=18mm,text width=22mm},
  trunk/.style={draw=gray!60,line width=0.4pt},
  arr/.style={->,>={Latex[length=1.2mm]},draw=gray!60,line width=0.4pt,shorten >=0.6mm}
]

\node[modality] (m1) at (0, 0.2)
  {Text / LLM\\\mdseries\scriptsize 4 papers\\\mdseries\scriptsize Sec.~\ref{sec:text-attacks}};
\node[family] (f1a) at (2.5, 1.2) {Trained generator};
\node[papers] (p1a) at (2.5,-0.4) {DiffusionAttacker;\\DART;\\Qiu et al.};
\node[family] (f1b) at (5.0, 1.2) {Off-the-shelf\\inference};
\node[papers] (p1b) at (5.0,-0.4) {L\"udke et al.};

\draw[trunk] (m1.east) -- (1.2, 0.2);
\draw[trunk] (1.2, 0.2) -- (1.2, 1.9);
\draw[trunk] (1.2, 1.9) -- (5.0, 1.9);
\draw[arr]   (2.5, 1.9) -- (f1a.north);
\draw[arr]   (5.0, 1.9) -- (f1b.north);

\node[modality] (m2) at (0,-3.5)
  {Image classifier\\\mdseries\scriptsize 18 papers\\\mdseries\scriptsize Sec.~\ref{sec:image-attacks}};
\node[family] (f2a) at (2.5,-2.5) {Latent\\perturbation};
\node[papers] (p2a) at (2.5,-4.1) {DiffAttack; ACA;\\Adv-Diffusion;\\DiffProtect;\\DiffAM};
\node[family] (f2b) at (5.0,-2.5) {Score guidance};
\node[papers] (p2b) at (5.0,-4.1) {AdvDiff;\\AdvDiffuser;\\Diff-PGD;\\ScoreAdv; SemDiff;\\NatADiff; APA};
\node[family] (f2c) at (7.5,-2.5) {Patch / physical};
\node[papers] (p2c) at (7.5,-4.1) {NaturalPatch;\\AdvLogo;\\DiffPatch /\\BadPatch; LSDM};
\node[family] (f2d) at (10.0,-2.5) {Purifier evasion};
\node[papers] (p2d) at (10.0,-4.1) {Kang et al.\\(DiffAttack\\vs DiffPure)};

\draw[trunk] (m2.east) -- (1.2,-3.5);
\draw[trunk] (1.2,-3.5) -- (1.2,-1.6);
\draw[trunk] (1.2,-1.6) -- (10.0,-1.6);
\draw[arr]   (2.5,-1.6) -- (f2a.north);
\draw[arr]   (5.0,-1.6) -- (f2b.north);
\draw[arr]   (7.5,-1.6) -- (f2c.north);
\draw[arr]   (10.0,-1.6) -- (f2d.north);

\node[modality] (m3) at (0,-7.2)
  {VLM\\\mdseries\scriptsize 10 papers\\\mdseries\scriptsize Sec.~\ref{sec:vlm-attacks}};
\node[family] (f3a) at (2.5,-6.2) {Score-guided};
\node[papers] (p3a) at (2.5,-7.8) {AdvDiffVLM;\\Xu et al.\\(cross-attention)};
\node[family] (f3b) at (5.0,-6.2) {Pipeline /\\renderer};
\node[papers] (p3b) at (5.0,-7.8) {HADES;\\MM-SafetyBench;\\IDEATOR;\\Visual-RolePlay;\\AttackVLM; InstructTA};
\node[family] (f3c) at (7.5,-6.2) {Diffusion-style\\pretrain};
\node[papers] (p3c) at (7.5,-7.8) {AnyAttack};
\node[family] (f3d) at (10.0,-6.2) {Diffusion as\\victim};
\node[papers] (p3d) at (10.0,-7.8) {AdvI2I\\(I2I diffusion)};

\draw[trunk] (m3.east) -- (1.2,-7.2);
\draw[trunk] (1.2,-7.2) -- (1.2,-5.3);
\draw[trunk] (1.2,-5.3) -- (10.0,-5.3);
\draw[arr]   (2.5,-5.3) -- (f3a.north);
\draw[arr]   (5.0,-5.3) -- (f3b.north);
\draw[arr]   (7.5,-5.3) -- (f3c.north);
\draw[arr]   (10.0,-5.3) -- (f3d.north);

\end{tikzpicture}

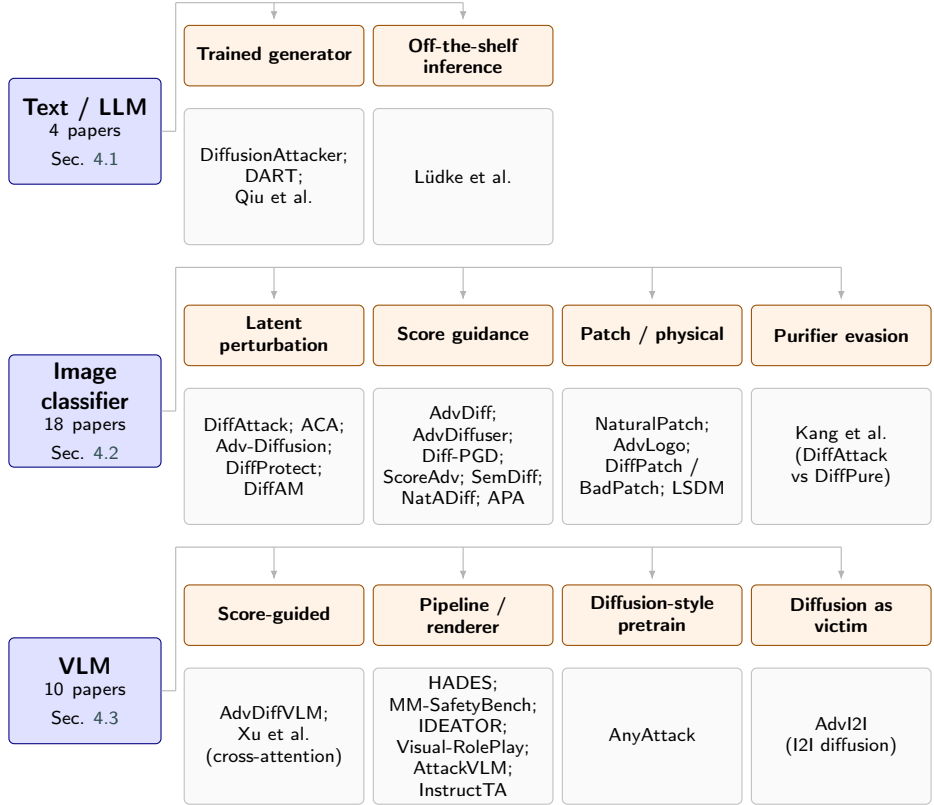
\captionof{figure}{Taxonomy of cataloged diffusion attacks: target
modality (rows) and diffusion role within each modality (columns).
Families~B, C, and F are treated separately in
Sections~\ref{sec:dllm-victim}, \ref{sec:defenses}, and~\ref{sec:baselines}.}
\label{fig:taxonomy}
\end{minipage}
\end{center}

Arranged by year (Figure~\ref{fig:timeline}), the catalog tells a different story. Family~D is the established cluster, accumulating steadily from 2023 through the May 2026 cutoff. Families~A and~B
emerged entirely in 2024--2025; the focal subjects of this survey are therefore also its newest entries. Twenty-two of the fifty papers (44\%) provide a public code release. Both figures are generated from the Catalog sheet of the companion repository
(Section~\ref{sec:data-synthesis}).

\begin{center}
\begin{minipage}{\linewidth}\centering
\begin{tikzpicture}
\begin{axis}[
    width=0.78\linewidth, height=5cm,
    xbar stacked,
    bar width=10pt,
    enlarge y limits=0.14,
    xmajorgrids=true, grid style={dashed,gray!25},
    xlabel={Cataloged papers},
    symbolic y coords={
      {A. Text/LLM},
      {B. dLLM victim},
      {C. Defenses},
      {E. VLM},
      {F. Baselines},
      {D. Vision}
    },
    ytick=data,
    nodes near coords,
    every node near coord/.append style={
        font=\scriptsize, color=black,
        /pgf/number format/.cd, fixed, precision=0
    },
    point meta=rawx,
    legend cell align=left,
    legend style={
        at={(0.5,-0.24)}, anchor=north, legend columns=3,
        font=\scriptsize, draw=none,
        /tikz/every even column/.append style={column sep=0.3cm}
    },
    xmin=0, xmax=18,
    xtick={0,2,4,6,8,10,12,14,16,18},
    every axis label/.append style={font=\footnotesize},
    tick label style={font=\scriptsize},
    yticklabel style={align=right}
]
\addplot+[fill=blue!25, draw=blue!55!black]
  coordinates {
    (0,{A. Text/LLM})
    (0,{B. dLLM victim})
    (1,{C. Defenses})
    (2,{E. VLM})
    (2,{F. Baselines})
    (5,{D. Vision})
  };
\addlegendentry{2023}
\addplot+[fill=blue!55, draw=blue!75!black]
  coordinates {
    (1,{A. Text/LLM})
    (0,{B. dLLM victim})
    (1,{C. Defenses})
    (5,{E. VLM})
    (7,{F. Baselines})
    (7,{D. Vision})
  };
\addlegendentry{2024}
\addplot+[fill=red!60, draw=red!80!black]
  coordinates {
    (3,{A. Text/LLM})
    (4,{B. dLLM victim})
    (2,{C. Defenses})
    (3,{E. VLM})
    (1,{F. Baselines})
    (6,{D. Vision})
  };
\addlegendentry{2025 (May 2026 cutoff)}
\end{axis}
\end{tikzpicture}

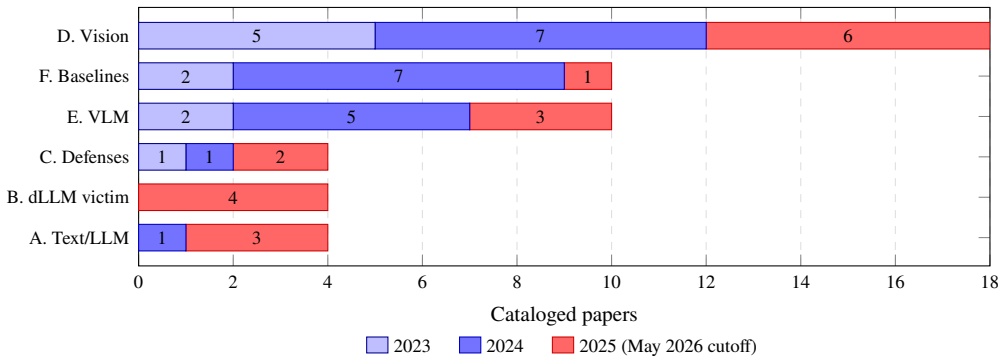
\captionof{figure}{Cataloged papers by family and year. Color
encodes year of publication; family ordering follows the survey's
focus rather than the alphabet.}
\label{fig:timeline}
\end{minipage}
\end{center}

\subsection{Diffusion Attacks on Text and Large Language Models}
\label{sec:text-attacks}
This is the smallest and most directly relevant cluster, comprising four
papers. Before turning to each in detail, it is worth pausing on a
question that the cluster as a whole answers in four genuinely different
ways: where in the adversarial pipeline does the diffusion process
actually live? Figure~\ref{fig:llm-cluster-philosophies} arranges the
four papers side-by-side along precisely this axis.

%
%
\begin{center}
\begin{minipage}{\linewidth}\centering
\begin{tikzpicture}[
  font=\sffamily\footnotesize,
  panel/.style={rectangle, rounded corners=3pt, draw=gray!40, fill=gray!2,
                inner sep=4pt, minimum width=70mm, minimum height=60mm},
  ptitle/.style={font=\sffamily\bfseries\small, anchor=north, text=blue!40!black},
  psub/.style={font=\sffamily\scriptsize, anchor=north, text=gray!50!black},
  diffregion/.style={rounded corners=5pt, fill=orange!8, draw=orange!50!black,
                     line width=0.4pt},
  diffreglabel/.style={font=\sffamily\bfseries\scriptsize, anchor=north,
                       text=orange!60!black, inner sep=2pt},
  victim/.style={rectangle, rounded corners=2pt, draw=blue!50!black, fill=blue!12,
                font=\sffamily\bfseries\scriptsize, align=center, inner sep=2pt,
                text width=12mm, minimum height=7mm},
  judge/.style={rectangle, rounded corners=2pt, draw=gray!55, fill=gray!10,
                font=\sffamily\bfseries\scriptsize, align=center, inner sep=2pt,
                text width=14mm, minimum height=7mm},
  output/.style={rectangle, rounded corners=2pt, draw=blue!50!black, fill=blue!12,
                font=\sffamily\bfseries\scriptsize, align=center, inner sep=2pt,
                text width=12mm, minimum height=9mm},
  tokmask/.style={font=\sffamily\tiny\itshape, text=gray!50, draw=gray!40,
                  fill=gray!10, rounded corners=1pt, inner sep=1pt,
                  minimum height=3.5mm, minimum width=4mm},
  tokfilled/.style={font=\sffamily\tiny, draw=orange!60!black,
                    fill=orange!15, rounded corners=1pt, inner sep=1pt,
                    minimum height=3.5mm, minimum width=4mm},
  parr/.style={->, >={Latex[length=1.2mm]}, draw=gray!70, line width=0.5pt,
               shorten >=0.4mm, shorten <=0.4mm},
  fbarr/.style={->, >={Latex[length=1.2mm]}, draw=orange!60!black,
                line width=0.5pt, dashed, shorten >=0.4mm, shorten <=0.4mm},
  parrlab/.style={font=\sffamily\scriptsize\itshape, text=gray!55!black},
  fbarrlab/.style={font=\sffamily\scriptsize\itshape, text=orange!60!black},
  leader/.style={line width=0.25pt, dotted},
  ptag/.style={font=\sffamily\scriptsize, align=center, text=blue!40!black,
              fill=blue!5, rounded corners=2pt, inner sep=3pt, text width=60mm}
]

\node[panel] (pn1) at (3.6, 1.5) {};
\node[ptitle] at (3.6, 4.45) {DiffusionAttacker};
\node[psub]   at (3.6, 4.10) {(Wang et al., EMNLP 2025)};

\node[diffregion, minimum width=26mm, minimum height=17mm] (reg1) at (1.65, 2.9) {};
\node[diffreglabel] at (reg1.north) {Continuous latent $z_t$};

\node[font=\scriptsize] at (0.75, 2.9) {$z_T$};
\node[font=\scriptsize] at (2.55, 2.9) {$z_0$};
\foreach \i in {0,...,4} {
  \pgfmathsetmacro{\opa}{0.20 + \i*0.18}
  \pgfmathsetmacro{\opaval}{int(\opa*100)}
  \fill[orange!\opaval!black, opacity=\opa] (1.00+\i*0.22, 3.10) circle (0.9pt);
  \fill[orange!\opaval!black, opacity=\opa] (1.00+\i*0.22, 2.70) circle (0.9pt);
}
\draw[->,>={Latex[length=0.8mm]}, draw=orange!60!black, line width=0.3pt]
  (1.00, 2.45) -- node[below=1pt, font=\sffamily\tiny\itshape, text=orange!60!black]
  {denoise} (2.25, 2.45);

\draw[parr] (reg1.east) -- node[parrlab, above=2pt] {LM\_head} (3.95, 2.9);
\draw[draw=gray!60, line width=0.3pt] (3.95, 2.60) rectangle (4.65, 3.30);
\foreach \i/\h in {0/0.12, 1/0.32, 2/0.50, 3/0.22, 4/0.30, 5/0.10} {
  \fill[blue!50!black] (4.02+\i*0.10, 2.65) rectangle ++(0.08, \h);
}
\node[anchor=north, font=\sffamily\tiny] at (4.30, 2.57) {logits};

\draw[parr] (4.70, 2.9) -- node[parrlab, above=2pt] {Gumbel} (5.55, 2.9);
\node[output] (out1) at (6.25, 2.9) {adv\\tokens};

\node[victim] (v1) at (3.6, 1.0) {Victim LLM};
\node[judge]  (j1) at (5.9, 1.0) {GPT-4 judge};
\draw[parr] (out1.south) -- ++(0, -0.6) -| (v1.north);
\draw[parr] (v1) -- (j1);

\draw[fbarr] (j1.south) -- ++(0, -0.40) -| (reg1.south);
\node[fbarrlab, font=\sffamily\tiny\itshape] at (3.6, 0.10)
  {attack loss $\to$ $\nabla$ Gumbel};

\node[ptag] at (3.6, -0.85)
  {Diffusion in \emph{continuous latent} $z_t$;\\
   Gumbel-Softmax bridges to differentiable discrete tokens};

\node[panel] (pn2) at (11.4, 1.5) {};
\node[ptitle] at (11.4, 4.45) {DART};
\node[psub]   at (11.4, 4.10) {(N\"other et al., AAAI 2025)};

\node[diffregion, minimum width=30mm, minimum height=20mm] (reg2) at (10.325, 2.85) {};
\node[diffreglabel] at (reg2.north) {Embedding space (T5)};

\fill[gray!50] (9.125, 2.20) circle (0.7pt);
\fill[gray!50] (11.375, 2.20) circle (0.7pt);
\fill[gray!50] (11.475, 3.45) circle (0.7pt);
\fill[gray!50] (9.175, 3.45) circle (0.7pt);

\draw[dashed, draw=orange!60!black, line width=0.4pt] (10.325, 2.85) circle (5mm);

\fill[blue!65!black] (10.175, 2.85) circle (2.0pt);
\node[anchor=east, font=\sffamily\tiny] (lblref) at (9.675, 2.30) {$x_{\text{ref}}$};
\draw[leader, draw=blue!65!black] (10.155, 2.84) -- (lblref.north east);

\fill[red!75!black] (10.475, 3.05) circle (2.0pt);
\node[anchor=west, font=\sffamily\tiny] (lblpert) at (11.025, 3.45) {$x_{\text{ref}}{+}\delta$};
\draw[leader, draw=red!75!black] (10.495, 3.06) -- (lblpert.south west);

\draw[->,>={Latex[length=1.2mm]}, draw=red!75!black, line width=0.7pt]
  (10.195, 2.87) -- (10.455, 3.03);

\node[font=\sffamily\tiny\itshape, text=orange!60!black]
  at (10.625, 2.55) {$\epsilon$};

\draw[parr] (reg2.east) -- node[parrlab, above=2pt] {decode} (12.675, 2.85);
\node[output] (out2) at (13.375, 2.85) {adv\\prompt};

\node[victim] (v2) at (11.125, 1.0) {Victim LLM};
\node[judge]  (j2) at (13.425, 1.0) {Tox.\ classif.};
\draw[parr] (out2.south) -- ++(0, -0.7) -| (v2.north);
\draw[parr] (v2) -- (j2);

\draw[fbarr] (j2.south) -- ++(0, -0.40) -| (reg2.south);
\node[fbarrlab, font=\sffamily\tiny\itshape] at (12.275, 0.10)
  {ASR signal (proximity-bounded)};

\node[ptag] at (11.4, -0.85)
  {Diffusion in \emph{embedding space};\\
   $\epsilon$-ball enforces proximity to reference prompt};

\node[panel] (pn3) at (3.6, -5.1) {};
\node[ptitle] at (3.6, -2.15) {L\"udke et al.};
\node[psub]   at (3.6, -2.50) {(arXiv 2025, LLaDA-8B)};

\node[diffregion, minimum width=60mm, minimum height=20mm] (reg3) at (3.6, -4.15) {};
\node[diffreglabel] at (reg3.north) {Joint $q(\mathbf{x},\mathbf{y})$ via pretrained DLLM};

\node[anchor=east, font=\sffamily\tiny\bfseries] at (0.95, -3.80) {$\mathbf{x}$:};
\foreach \i in {0,...,4} {
  \node[tokmask] at (1.10+\i*0.40, -3.80) {[M]};
}
\draw[->,>={Latex[length=0.8mm]}, draw=orange!60!black, line width=0.3pt]
  (3.10, -3.80) -- node[font=\sffamily\tiny\itshape, text=orange!60!black, above=-1pt]
  {sample} (3.80, -3.80);
\node[tokfilled] at (4.10, -3.80) {Tell};
\node[tokfilled] at (4.50, -3.80) {me};
\node[tokfilled] at (4.90, -3.80) {how};
\node[tokfilled] at (5.30, -3.80) {to};
\node[tokfilled] at (5.70, -3.80) {\dots};

\node[anchor=east, font=\sffamily\tiny\bfseries] at (0.95, -4.50) {$\mathbf{y}$:};
\node[tokfilled] at (1.10, -4.50) {Sure};
\node[tokfilled] at (1.50, -4.50) {,};
\node[tokfilled] at (1.90, -4.50) {here};
\node[tokfilled] at (2.30, -4.50) {is};
\node[tokfilled] at (2.70, -4.50) {the};
\node[tokfilled] at (3.10, -4.50) {step};
\node[tokfilled] at (3.50, -4.50) {by};
\node[tokfilled] at (3.90, -4.50) {step};
\node[tokfilled] at (4.30, -4.50) {\dots};
\node[anchor=west, font=\sffamily\tiny\itshape, text=orange!60!black]
  at (4.65, -4.50) {(fixed)};

\node[anchor=center, font=\sffamily\scriptsize\itshape, text=orange!60!black]
  at (3.6, -4.90) {$\circlearrowleft$ 2000 random restarts \quad no training};

\node[victim] (v3) at (3.6,  -6.35) {Victim LLM};
\node[judge]  (j3) at (5.9,  -6.35) {JBB + SR};
\draw[parr] (5.00, -4.00) -- ++(0, -0.4) -| (v3.north);
\draw[parr] (v3) -- (j3);

\node[ptag] at (3.6, -7.65)
  {Diffusion in \emph{masked token grid}; inpaint $\mathbf{x}{\sim}q(\mathbf{x}|\mathbf{y})$;\\
   pretrained DLLM, no training, restart-based};

\node[panel] (pn4) at (11.4, -5.1) {};
\node[ptitle] at (11.4, -2.15) {Qiu et al.};
\node[psub]   at (11.4, -2.50) {(SPIE ICCAID 2023)};

\node[diffregion, minimum width=60mm, minimum height=22mm] (reg4) at (11.4, -4.15) {};
\node[diffreglabel] at (reg4.north) {Diffusion w/ dual-objective training};

\node[anchor=center, font=\sffamily\tiny, draw=gray!50, fill=white,
      rounded corners=1pt, inner sep=2pt, text width=10mm, align=center]
  (xorig) at (9.20, -4.15) {original\\text $x$};

\node[anchor=center, font=\sffamily\bfseries\scriptsize, draw=orange!60!black,
      fill=orange!15, rounded corners=2pt, inner sep=2pt, text width=12mm,
      minimum height=8mm, align=center]
  (diff4) at (10.80, -4.15) {Diffusion\\model};

\draw[parr] (xorig) -- (diff4);

\node[anchor=center, font=\sffamily\tiny, draw=blue!50!black, fill=blue!12,
      rounded corners=1pt, inner sep=2pt, text width=10mm, align=center]
  (xadv) at (12.40, -4.15) {adv text\\$x^\prime$};

\draw[parr] (diff4) -- (xadv);

\node[anchor=center, font=\sffamily\tiny, text=orange!60!black,
      align=center, text width=12mm]
  (semobj) at (13.75, -3.75) {semantic\\sim($x,x^\prime$)};
\node[anchor=center, font=\sffamily\tiny, text=orange!60!black,
      align=center, text width=12mm]
  (advobj) at (13.75, -4.55) {surrogate\\classifier\\(adv loss)};

\draw[->,>={Latex[length=0.8mm]}, draw=orange!60!black, line width=0.3pt, dashed]
  (xadv.north east) to[bend left=12] (semobj.west);
\draw[->,>={Latex[length=0.8mm]}, draw=orange!60!black, line width=0.3pt, dashed]
  (xadv.south east) to[bend right=12] (advobj.west);

\draw[fbarr] (advobj.south) to[bend left=45]
  node[pos=0.55, below=0pt, fbarrlab, font=\sffamily\tiny\itshape]
  {dual loss}
  (diff4.south);

\node[victim] (v4) at (10.80, -6.35) {\textbf{Target} classif.};
\node[judge]  (j4) at (13.10, -6.35) {ASR};
\draw[parr] (xadv.south) -- ++(0, -0.5) -|
  node[pos=0.25, below=0pt, font=\sffamily\tiny\itshape, text=blue!40!black]
  {black-box transfer} (v4.north);
\draw[parr] (v4) -- (j4);

\node[ptag] at (11.4, -7.65)
  {Diffusion trained with \emph{dual objectives}: semantic +\\
   adversarial; black-box transfer to target classifier};

\end{tikzpicture}

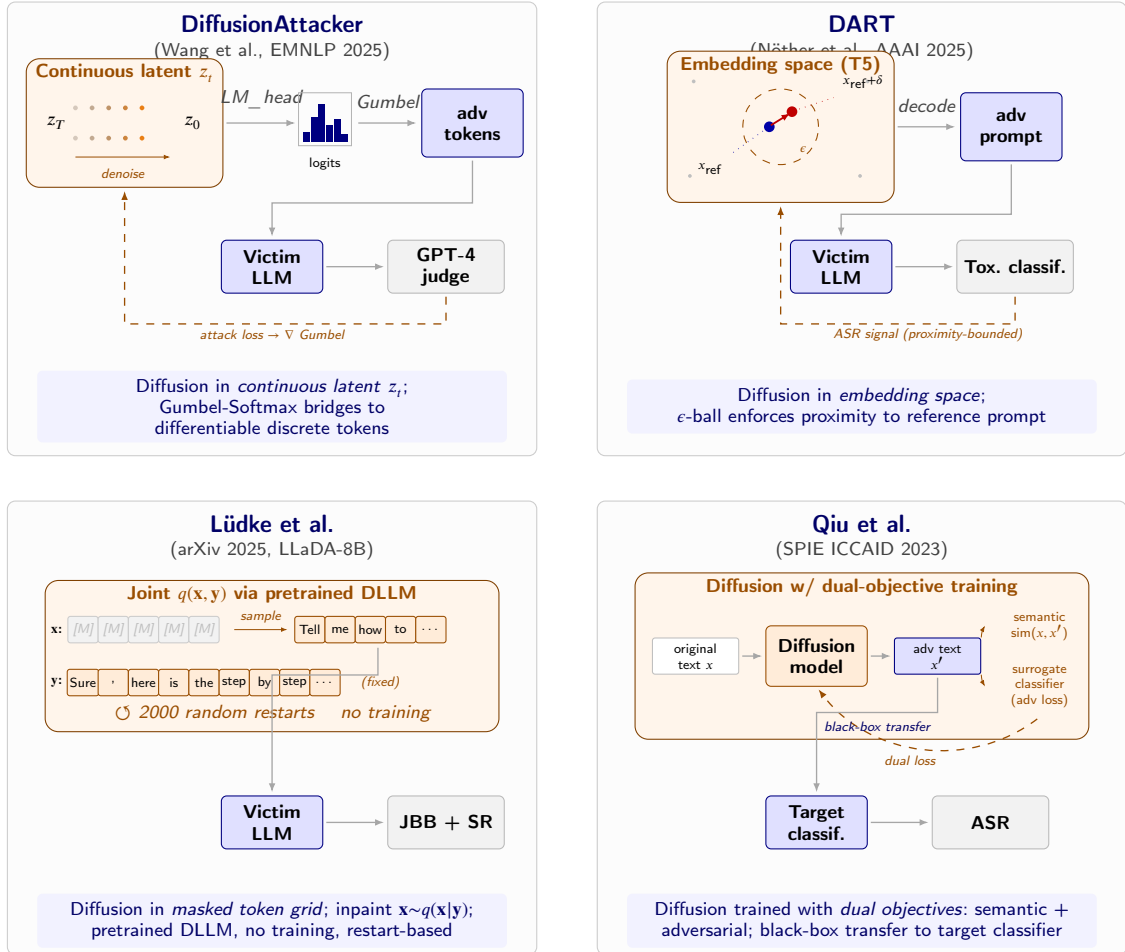
\captionof{figure}{Mechanistic comparison of the four Family~A papers. Section~\ref{sec:text-attacks} describes each
panel in detail.}
\label{fig:llm-cluster-philosophies}
\end{minipage}
\end{center}

DiffusionAttacker (top-left panel) places the diffusion process in a
continuous latent space: the generator denoises a noise vector $z_T$
into $z_0$ over several steps, the LM\_head maps the resulting latent
into a logit distribution over the vocabulary, and Gumbel-Softmax
samples discrete tokens in a way that remains differentiable, so the
attack loss can flow back through the sampler at training time.
DART (top-right panel) instead operates in the embedding space of a T5
backbone: a reference prompt's embedding $x_{\text{ref}}$ is perturbed
within a hard $\epsilon$-ball, and the resulting embedding
$x_{\text{ref}}{+}\delta$ is decoded back to natural language. The
dashed circle is the central commitment of the paper, not decoration.
L\"udke et al.\ (bottom-left panel) work directly in masked-token space
and dispense with training entirely: the harmful response $\mathbf{y}$
is held fixed, the prompt $\mathbf{x}$ starts fully masked, and a
pretrained masked diffusion language model samples $\mathbf{x}$
conditioned on $\mathbf{y}$ across roughly two thousand random
restarts. The panel has no feedback arrow because there is no training
to feed back into. Qiu et al.\ (bottom-right panel) trains a diffusion
model with two objectives simultaneously, a semantic-preservation term
and an adversarial loss against a surrogate classifier, and relies on
black-box transfer to attack the actual target classifier, which the
diffusion model never sees during training.

The contrast the figure makes visible is that the four papers do not
share a single methodology. They share a high-level commitment to using
diffusion somewhere in the adversarial pipeline, but below that they
pursue four genuinely different mechanisms: continuous latent denoising
with differentiable sampling, embedding-space perturbation under a
proximity constraint, inference-time inpainting on a pretrained masked
DLLM, and dual-objective embedding-space training for transfer attack.
This heterogeneity is worth keeping in mind when reading the cluster
critique in Section~\ref{sec:cluster-critique-llm}: many of the
weaknesses we identify are weaknesses of specific mechanisms,
not of the cluster as a whole. We now take each paper in turn.

\subsubsection{DiffusionAttacker}
\label{sec:diffusionattacker}
DiffusionAttacker~\citep{wang2024diffusionattacker} reformulates
jailbreaking as sequence-to-sequence rewriting. The attacker is a
continuous text-diffusion generator that conditions on a source harmful
instruction and produces a rewritten prompt through iterative denoising
in a continuous latent space; the attack loss, computed against a target
LLM, is backpropagated through the sampler via Gumbel-Softmax
relaxation, which makes the otherwise non-differentiable
discrete-token sampling step at the end of the pipeline differentiable.
The paper's positioning claim is that the seq2seq formulation overcomes
a structural limitation of autoregressive attackers such as
AdvPrompter~\citep{paulus2024advprompter}: an autoregressive attacker
can only append tokens and cannot revise what it has already emitted,
whereas a seq2seq diffuser can re-edit any position during denoising.

Experiments run on AdvBench and HarmBench against four open-weight
aligned LLMs. The reported white-box attack success rates, given as the
prefix-match / GPT-judge pair ASR\textsubscript{prefix}/ASR\textsubscript{GPT},
reach 90/74\% on Llama-3-8B, 93/79\% on Mistral-7B, 91/77\% on Vicuna-7B,
and 88/71\% on Alpaca-7B+Safe-RLHF. The paper also reports transfer to
closed frontier models, though through a particular framing: rather than
apply the trained diffusion attacker directly, the authors use
DiffusionAttacker as a rewriter that feeds existing black-box methods.
Under that framing, transfer numbers are 56/49\% on GPT-4o and 33/21\%
on Claude-3.5; Gemini is not evaluated.

The contribution is real but bounded. The paper demonstrates that a
trained seq2seq diffusion attacker is feasible, and the Gumbel-Softmax
differentiation through the sampler is a useful methodological addition
to the discrete-attack literature. The white-box dependency at training
time is substantial, however: the attack loss is backpropagated through
each target LLM's own parameters, so the white-box ASR reported above is
partly a statement about each target's gradient surface, and the
substantial dropoff from white-box (90/74\% on Llama-3-8B) to
black-box-transfer numbers (56/49\% on GPT-4o, 33/21\% on Claude-3.5)
is consistent with this reading. A second concern is structural to the
diversity metric: stochastic diffusion samplers are by construction
more diverse than greedy autoregressive decoders, so the reported
diversity gain over AdvPrompter is partly an artifact of decoding
strategy, and whether the additional diversity translates into harm a
defender would care about is a separate question.

\subsubsection{DART}
\label{sec:dart}
DART~\citep{noether2025dart} poses red teaming as constrained search.
Every discovered prompt must lie within an L2 ball of a reference prompt
in embedding space, anchoring the attack to a specific topic, writing
style, or category of harmful behavior. The mechanism is a learned
perturbation model trained with diffusion-style noising: it produces
bounded perturbations of reference embeddings that maximize a learned
harmfulness reward, and the perturbed embedding is decoded back to text
via a fixed decoder. The proximity constraint differentiates DART from
gibberish-suffix attacks such as GCG~\citep{zou2023universal} and from
genetic-algorithm refinements such as AutoDAN~\citep{liu2024autodan},
neither of which guarantees semantic proximity to a reference.

The reported targets are three aligned LLMs of increasing safety
(gpt2-alpaca, Vicuna-7B, and Llama-2-7B-chat-hf), and the headline result
is a Pareto frontier of (cosine-similarity-to-reference, ASR) that
dominates each comparison baseline (unmodified prompts, RL fine-tuning,
zero-shot prompting, few-shot prompting, and
FLIRT~\citep{mehrabi2023flirt}) under tight proximity budgets
($\epsilon=0.1$ and $\epsilon=0.5$). FLIRT does achieve higher
unconstrained toxicity than DART, but only by drifting away from the
reference; once a proximity budget is imposed, DART discovers more toxic
prompts within the budget than any baseline. The authors further show
that established autoregressive architectures perform poorly under the
proximity constraint, supporting the case for a generator that can move
in arbitrary directions in embedding space rather than only forward.

Of the four Family~A papers, DART is the methodologically most novel: it
reframes the problem (anchored audit of a topic or harmful-behavior
class) rather than trying to outperform GCG on the conventional
ASR-on-AdvBench benchmark. The principal caveat is terminological:
DART's use of the word ``diffusion'' refers to a learned noise model
trained with diffusion-style objectives, not to a full multi-step
denoising-diffusion sampler with an iterative reverse process. This
does not detract from the contribution but shapes how the work should
be situated: DART belongs to the broader family of learned-perturbation
attackers that take inspiration from diffusion training, rather than to
the class of true denoising-diffusion attackers. A second caveat,
shared with DiffusionAttacker, is that transfer to frontier closed
models is not reported.

\subsubsection{Diffusion LLMs as Natural Adversaries}
\label{sec:ludke}
\citet{ludke2025diffusion} use pretrained diffusion LLMs as off-the-shelf
adversarial generators against any target LLM. The key observation is
that masked diffusion language models such as LLaDA~\citep{nie2025llada}
jointly model prompt-response pairs, $p(\text{prompt}, \text{response})$.
Fixing the response to a desired harmful completion and inpainting the
prompt under this joint distribution therefore yields adversarial
prompts in a fully training-free manner. The procedure is conceptually
simple: choose a target completion, instantiate the dLLM, mask the
prompt portion, and run a conditional inpainting pass with a small
number of parallel samples. A probabilistic analysis in the paper shows
that under a fidelity assumption on the dLLM, namely that it accurately
models $p(\text{prompt}\mid\text{response})$ in the relevant region of
the joint distribution, a small number of conditional samples suffices
to recover high-reward prompts.

Empirically, the reported ASR figures span a wide range. With 2{,}000
random restarts per target and no per-target training, the attack
achieves 100\% on Phi-4-Mini, Qwen-2.5-7B, and Llama-3-8B; 91\% on
LAT-hardened Llama-3-8B; 93\% on Circuit Breakers; 99\% on Gemma-3-1B;
and 53\% on ChatGPT-5 over 100 attack attempts. The authors also report
that the generated prompts are low-perplexity (and so plausibly evade
perplexity-based filters) and diverse across restarts. All evaluation
is on JailbreakBench / StrongREJECT.

This is the first paper to operationalize a true masked-diffusion
sampler as a zero-cost attacker against autoregressive targets, and it
is the cleanest realization of the ``diffusion as natural adversary''
idea: no per-instance optimization, no attack-specific training, and no
white-box access to the target. The principal limitation is that the
fidelity assumption is doing real work. If the pretrained dLLM has not
been exposed to the specific harmful intents of interest during
pretraining, which is plausible because pretraining data are typically
filtered to remove harmful content, then
$p(\text{prompt}\mid\text{response})$ for those completions may be
unreliable. The paper is recent (October 2025) and at the time of
writing the authors had not released code, which limits independent
replication. The practical implication for follow-on work is clear: any
new diffusion-based attacker on LLMs must explicitly differentiate from
response-conditional inpainting on a pretrained dLLM, because this
baseline is now nearly free.

\subsubsection{Diffusion-Based Adversarial Attack on NLP Classifiers}
\label{sec:qiu}
\citet{qiu2024diffusion} is the earliest entry in this cluster,
predating the LLM-jailbreak wave by roughly a year. The targets are
conventional text-classification models (sentiment, topic, and related
discriminative tasks) rather than chat LLMs. The attacker is an
embedding-space diffusion model trained jointly with a substitute
classifier, and the training objective combines two terms: a
semantic-preservation term that keeps the perturbed embedding decoding
back to text close in meaning to the input, and an adversarial term
that drives the substitute classifier toward a wrong label. The
black-box transfer assumption is standard: a victim classifier of
similar architecture but different weights will be fooled by adversarial
examples crafted against the substitute. The paper reports that the
proposed method successfully generates adversarial texts that fool
text-classification targets, but the publicly available abstract does
not give a specific ASR figure or improvement-over-baseline number. The
paper appears in a comparatively low-visibility venue (the SPIE ICCAID
2023 proceedings), which limits the methodological detail accessible
without institutional access.

We include Qiu et al.\ as the historical reference point for
``diffusion as adversarial text generator,'' and because the
text-classifier attack thread, as opposed to the LLM-jailbreak thread,
remains essentially uncultivated. A clean modern follow-up using a true
discrete-diffusion sampler against text-classification targets has not
been published, and we flag it as one of the more tractable
opportunities in Section~\ref{sec:implications}.

\subsubsection{Cluster Critique}
\label{sec:cluster-critique-llm}
Read together, the four papers expose several recurring weaknesses. The
first is a pervasive white-box dependency: three of the four train
against white-box gradients of the target model. Direct training-free
attacks on closed frontier models are reported only by
\citet{ludke2025diffusion}, and even there the assumption that
pretrained dLLMs have learned a good $p(\text{prompt}\mid\text{response})$
for harmful completions is an empirical question that bears further
scrutiny. DiffusionAttacker does report transfer to GPT-4o (56/49\%
ASR\textsubscript{prefix}/ASR\textsubscript{GPT}) and Claude-3.5
(33/21\%), but these numbers reflect the use of DiffusionAttacker as a
rewriter for existing black-box methods rather than direct application
of the trained diffusion sampler to the closed target.

Second, no paper combines a true discrete diffusion sampler
(D3PM, SEDD, MDLM) with an explicit harmfulness reward and a
HarmBench-scale evaluation; each picks one or two of these components
but not all three.

Third, defense-evasion is under-evaluated. The papers claim low
perplexity for generated prompts but none benchmarks the attacks against
modern input filters such as Llama-Guard~\citep{inan2023llamaguard} or
SmoothLLM~\citep{robey2023smoothllm}, and reported low-perplexity
numbers are necessary but not sufficient evidence of filter evasion.

Fourth, multi-turn attacks are absent. All four papers produce
single-shot adversarial prompts, even though diffusion's natural
strength, joint modeling of long sequences with bidirectional
conditioning, is precisely what one would want for adversarial
multi-turn dialogue.

Fifth, constrained or topic-specific red teaming is represented by a
single paper~\citep{noether2025dart}. Combining DART's proximity
constraint with a true discrete-diffusion sampler is an obvious
unfilled cell of the methodological matrix. Table~\ref{tab:summary-text-attacks} summarizes the four Family~A
papers.

{\small
\begin{longtable}{P{1.8cm} P{1.3cm} P{2.2cm} P{2.5cm} P{0.8cm} P{4.0cm} P{0.5cm}}
\caption{Diffusion-based adversarial attacks on text and language
models (Family~A). Threat: \emph{WB train} indicates white-box gradients
used during training; \emph{BB} indicates black-box attack or transfer;
\emph{GB} indicates grey-box access (surrogate or encoder-only).
\cmark{}: public code released.}
\label{tab:summary-text-attacks}\\
\toprule
Reference & Venue / yr & Diffusion role & Target & Threat & Dataset / key reported finding & Code \\
\midrule
\endfirsthead
\multicolumn{7}{c}{\tablename\ \thetable{} -- continued from previous page} \\
\toprule
Reference & Venue / yr & Diffusion role & Target & Threat & Dataset / key reported finding & Code \\
\midrule
\endhead
\midrule
\multicolumn{7}{r}{continued on next page} \\
\endfoot
\bottomrule
\endlastfoot
\citet{wang2024diffusionattacker} & EMNLP 2025 & Trained generator (seq2seq text diffusion) & Llama-3, Vicuna, Mistral, Alpaca+Safe-RLHF; transfer to GPT-4o, Claude-3.5 & WB train; BB transfer & AdvBench, HarmBench. ASR\textsubscript{prefix}/ASR\textsubscript{GPT} 90/74 on Llama-3-8B, 93/79 Mistral-7B, 91/77 Vicuna-7B, 88/71 Alpaca-7B+Safe-RLHF. BB transfer (as rewriter): 56/49 GPT-4o, 33/21 Claude-3.5. & \xmark \\
\citet{noether2025dart} & AAAI 2025 & Trained perturbation (diffusion-style noising) & gpt2-alpaca, Vicuna-7B, Llama-2-7B-chat-hf & BB & Reference-prompt sets. Pareto frontier of (cosine-sim, ASR) dominates baselines (unmodified, RL FT, zero/few-shot, FLIRT) under $\epsilon{=}0.1$ and $\epsilon{=}0.5$. & \xmark \\
\citet{ludke2025diffusion} & arXiv 2025 & Off-the-shelf inference (pretrained dLLM as $p(\text{prompt}\mid\text{response})$) & Phi-4-Mini, Qwen-2.5-7B, Llama-3-8B (incl.\ LAT, Circuit Breakers), Gemma-3-1B, ChatGPT-5 & BB & JailbreakBench / StrongREJECT (2{,}000 restarts). 100\% on Phi-4-Mini, Qwen-2.5-7B, Llama-3-8B; 91\% LAT Llama-3-8B; 93\% Circuit Breakers; 99\% Gemma-3-1B; 53\% ChatGPT-5 (100 attempts). & \xmark \\
\citet{qiu2024diffusion} & SPIE 2024 & Trained generator (embedding diffusion + substitute classifier) & Text classifiers (sentiment, topic) & GB & Standard text-classification benchmarks. First demonstration of diffusion-based adversarial text against NLP classifiers; specific ASR not reported in abstract. & \xmark \\
\end{longtable}
}

\subsection{Diffusion Attacks on Image Classifiers}
\label{sec:image-attacks}
The image-domain literature is substantially larger than the text-domain literature and provides the methodological vocabulary that text-side authors are now porting. We organize it into four families.

\subsubsection{Latent-Space Optimization through DDIM Inversion}
\citet{chen2024diffattack} introduce DiffAttack, the first paper of any kind to use a diffusion model for adversarial example crafting. The recipe is to DDIM-invert a clean image to its latent, and then back-propagate a classifier loss through the reverse stochastic differential equation (SDE) to update the latent. Two regularization terms, one that deviates cross-attention from the clean prediction and one that preserves self-attention, balance attack effectiveness against perceptual structure. \citet{chen2023content} extend this with manifold-constrained projected gradient descent (PGD) on Stable Diffusion latents and report substantial transferability gains over prior unrestricted-adversarial-example methods. \citet{liu2024advdiffusion} and \citet{sun2024diffam} apply the same recipe to face recognition with identity-preserving inpainting and makeup transfer, respectively. The text analogue of DDIM-inversion is invertible embedding-space diffusion: encode a clean prompt to embeddings, run a discrete-diffusion forward to a noisy latent, then optimize a judge loss while denoising back. DiffusionAttacker is the closest analogue, but it treats the diffuser as a parametric attacker rather than as a frozen prior.

\subsubsection{Score / Classifier Guidance during Reverse Sampling}
\citet{dai2024advdiff} introduce AdvDiff, which injects an adversarial guidance term directly into the reverse SDE: the class-likelihood guidance term during reverse sampling and a noise-sampling adversarial guidance term jointly produce realistic samples whose adversariality is internalized in the score. \citet{chen2023advdiffuser} perturb the predicted clean sample at each denoising step with class activation map (CAM) masking to preserve salient regions, achieving near-100\% ASR against RobustBench leaders on CIFAR-10, CelebA, and ImageNet. \citet{xue2023diffpgd} wrap standard PGD with a per-step diffusion projection that re-pulls perturbed images onto the natural-image manifold. \citet{huang2025scoreadv} reweight the denoising posterior with a classifier-score adversarial term at every DDIM step in a fully training-free manner. \citet{collins2025natadiff} target the decision-boundary intersection of the true and adversarial classes, producing samples that resemble naturally occurring mistakes. \citet{dai2025semdiff} optimize multiple semantic attributes in the deep semantic latent space (h-space) of a pretrained diffusion model, and \citet{jiang2025apa} bring Direct Preference Optimization (DPO) machinery to adversarial diffusion through a two-stage preference-alignment framework that decouples conflicting preferences. The text analogue is a harmfulness-classifier guidance term during the reverse process of a discrete diffusion sampler. No published paper has implemented this recipe directly for text.

\subsubsection{Diffusion as a Generator of Physical or Patch Attacks}
\citet{chen2023naturalpatch}, \citet{miao2024advlogo}, \citet{wang2024diffpatch} (renamed BadPatch in the most recent revision), and \citet{wang2025lsdm} use latent diffusion to generate adversarial patches and global perturbations against object detectors (YOLO family, Faster R-CNN, DETR). The recipes vary: latent perturbation under detector loss with expectation over transformations (EOT) for printability~\citep{chen2023naturalpatch}; Fourier-domain perturbation of latent and unconditional embeddings at the last diffusion timestep~\citep{miao2024advlogo}; null-text inversion with masked latent attack to preserve user-customizable styles~\citep{wang2024diffpatch}; and per-step latent perturbation conditioned on the input image for global non-patch detector evasion~\citep{wang2025lsdm}. The text analogue, by analogy, is wrapping a harmful query inside an in-context-learning template, a roleplay frame, or a persona; some VLM-side papers (HADES, Visual-RolePlay) realize this analogue partially through diffusion-rendered scene or character imagery.

\subsubsection{Diffusion-Purification Evasion}
DiffPure~\citep{nie2022diffusion}, a diffusion-based input purifier that runs a short forward-noising-then-denoising pass with a pretrained diffusion model to project adversarial inputs back toward the natural-image manifold, has become the canonical diffusion-based defense in the image domain. \citet{kang2024diffattack} (note: a different DiffAttack than the one of \citealp{chen2024diffattack}) and the broader adaptive-attack-on-DiffPure line demonstrate that diffusion-based defenses can be broken by gradient-aware adaptive attacks that back-propagate through the full denoising chain, with segment-wise back-propagation used to bound memory. The text analogue is straightforward and unaddressed: no published adaptive attack has yet broken DiffuseDef \citep{li2025diffusedef} or MaskPure \citep{gietz2024maskpure}, the two main diffusion-based text purifiers, nor CoDefend \citep{zhu2025codefend}, the corresponding multimodal-side defense. We return to this opportunity in Section~\ref{sec:implications}.

\subsubsection{Other Image-Side Work}
Two cataloged papers fit the four families above but extend rather than redefine them. \citet{liu2023diffprotect} use a diffusion autoencoder to generate semantically meaningful expression-level perturbations for face-recognition privacy protection, sitting at the intersection of latent-space optimization and the face-attack subcluster. \citet{li2025diffattackx} extend the original DiffAttack with bi-level routing attention and focal-loss reweighting for small features, an incremental variant of the latent-space-optimization family. We cite these works for catalog completeness but do not regard either as defining a new methodological family. Table~\ref{tab:summary-image} summarizes the eighteen image-classifier papers discussed in this subsection.

{\small
\begin{longtable}{P{1.8cm} P{1.3cm} P{2.2cm} P{1.6cm} P{0.6cm} P{4.5cm} P{0.5cm}}
\caption{Diffusion-based adversarial attacks on image classifiers, face-recognition models, and object detectors (Family D). Threat column: WB white-box, GB grey-box (surrogate or encoder-only access). \cmark{}: public code released.}
\label{tab:summary-image}\\
\toprule
Reference & Venue / yr & Diffusion role & Target & Threat & Dataset / key reported finding & Code \\
\midrule
\endfirsthead
\multicolumn{7}{c}{\tablename\ \thetable{} -- continued from previous page} \\
\toprule
Reference & Venue / yr & Diffusion role & Target & Threat & Dataset / key reported finding & Code \\
\midrule
\endhead
\midrule
\multicolumn{7}{r}{continued on next page} \\
\endfoot
\bottomrule
\endlastfoot
\citet{chen2024diffattack} & TPAMI 2024 & Frozen, latent perturbation (DDIM-inversion) & ResNet, ViT, Swin, ConvNeXt & WB & ImageNet. First diffusion-based AE crafting; high transferability across CNNs and ViTs. & \cmark \\
\citet{chen2023content} & NeurIPS 2023 & Frozen, latent PGD (Stable Diffusion latent) & ImageNet classifiers, including defended & GB & ImageNet. +13--50\% ASR over prior unrestricted-AE methods on undefended; +17--48\% on defended. & \cmark \\
\citet{liu2024advdiffusion} & AAAI 2024 & Frozen, latent inpainting (identity-preserving) & ArcFace, CosFace, FaceNet, IRSE, MobileFace & GB & FFHQ, CelebA-HQ. Identity-preserving face-recognition attack with strong cross-backbone transfer. & \cmark \\
\citet{liu2023diffprotect} & arXiv 2023 & Diffusion autoencoder (semantic latent) & ArcFace and similar FR backbones & GB & CelebA-HQ, FFHQ. Expression-level perturbations more natural than pixel noise. & \cmark \\
\citet{sun2024diffam} & CVPR 2024 & Frozen, conditional latent diffusion (makeup style) & ArcFace, IRSE, MobileFace & GB & CelebA-HQ, FFHQ. Adversarial makeup transfer combining CLIP and FR losses. & \xmark \\
\citet{dai2024advdiff} & ECCV 2024 & Frozen, score guidance & ResNet-50, WideResNet, Inception-v3 & WB & MNIST, ImageNet. Class-likelihood + noise-sampling adversarial guidance internalized in the score. & \cmark \\
\citet{chen2023advdiffuser} & ICCV 2023 & Frozen, score guidance ($x_0$-prediction perturb.) & RobustBench top defenses (Salman'20, Wang'23) & WB & CIFAR-10, CelebA, ImageNet. LPIPS $\sim$6$\times$ lower vs GA-Attack; FID 2--3$\times$ lower; near-100\% ASR. & \cmark \\
\citet{xue2023diffpgd} & NeurIPS 2023 & Frozen, score guidance + diffusion projection & ResNet, ViT, DiffPure-defended & WB & ImageNet. PGD wrapped with per-step diffusion projection re-pulls perturbations onto the natural-image manifold. & \cmark \\
\citet{huang2025scoreadv} & arXiv 2025 & Frozen, posterior reweighting & ResNet, ViT, ArcFace; ten target models & WB/BB & ImageNet, CelebA. SOTA ASR with image quality preserved; robust under DiffPure / NRP / JPEG defenses. & \xmark \\
\citet{dai2025semdiff} & arXiv 2025 & Frozen, semantic-latent (h-space) search & Standard classifiers + CLIP zero-shot & GB & CelebA-HQ, AFHQ, ImageNet. Multi-attribute joint optimization in the deep semantic latent space. & \xmark \\
\citet{collins2025natadiff} & arXiv 2025 & Classifier-guided + time-travel sampling & ResNet, ViT, ConvNeXt & GB & ImageNet (and ImageNet-A-style natural-AE distribution). Generated samples resemble naturally occurring mistakes. & \xmark \\
\citet{jiang2025apa} & arXiv 2025 & Preference-fine-tuned diffusion (DPO-style) & ResNet, ViT, robust models & WB & ImageNet. Two-stage preference-alignment framework decoupling visual quality and attack effectiveness. & \xmark \\
\citet{chen2023naturalpatch} & arXiv 2023 & Frozen, latent diffusion + EOT & YOLO v2/v3/v5, Faster R-CNN & WB & Object-detection benchmarks. Printable, semi-natural adversarial patches via diffusion. & \xmark \\
\citet{miao2024advlogo} & arXiv 2024 & Frozen, latent + Fourier perturbation & YOLO v2/v3/v4/v4-tiny/v5, Faster R-CNN, SSD & WB & Detection benchmarks. Logo-style patches mimicking brand artwork. & \cmark \\
\citet{wang2024diffpatch} & arXiv 2024 & Frozen, null-text inversion + masked latent & Person detectors (YOLOv5/v7) & WB & AdvT-shirt-1K (released). Customizable T-shirt-style adversarial patches. & \cmark \\
\citet{wang2025lsdm} & Neuro-computing 2025 & Frozen, per-step latent perturbation & YOLOv5, Faster R-CNN, DETR & GB & COCO/VOC. Average detection mAP -1.52\% with image-quality +1.71\% over baselines. & \xmark \\
\citet{li2025diffattackx} & Applied Intelligence 2025 & Frozen, latent perturbation (extension of DiffAttack) & ResNet, ViT, Swin & GB & ImageNet-compatible. Bi-level routing attention + focal-loss reweighting for small features. & \xmark \\
\citet{kang2024diffattack} & NeurIPS 2024 & Adaptive attack through diffusion purifier & DiffPure-defended ResNet/WRN; score-based purifiers & WB & CIFAR-10, ImageNet. -20\% robust accuracy on CIFAR-10 ($\epsilon=8/255$); -10\% on ImageNet ($\epsilon=4/255$). & \xmark \\
\end{longtable}
}

\subsection{Diffusion Attacks on Vision-Language Models}
\label{sec:vlm-attacks}
The VLM literature is the second-largest cluster (ten papers in our catalog) but is dominated by pipeline-only uses of diffusion. We separate the cluster by diffusion role.

\subsubsection{Score-Guided Diffusion Attacks on VLMs}
\citet{guo2024advdiffvlm} introduce AdvDiffVLM, the first true score-guided diffusion attack against commercial VLMs. The approach uses Adaptive Ensemble Gradient Estimation to modify the score during reverse sampling and CAM-guided masks to disperse adversarial signal spatially. The targets include several open-weight VLMs as well as commercial systems including GPT-4V. \citet{xu2024transferable} perturb the cross-attention activations of Stable Diffusion during sampling and report strong transferability across CLIP, ALBEF, TCL, and BLIP for retrieval and visual question answering (VQA) tasks; their adversarial generation is on a vision-language pretraining axis rather than a chat-VLM jailbreak axis.

\subsubsection{Pipeline-Only Diffusion Attacks on VLMs}
A larger group of papers uses diffusion as a black-box content renderer inside a larger attack pipeline. \citet{li2024hades} introduce HADES, which hides harmful intent in typography, amplifies it with a Stable-Diffusion-rendered scene, and adds a pixel-space adversarial overlay; the gradient flows only through the overlay, not through the diffusion sampler. \citet{liu2024mmsafetybench} build MM-SafetyBench, a benchmark that uses GPT-4 keyword extraction with Stable Diffusion rendering and typographic overlays to construct 5{,}040 image-text pairs probing thirteen unsafe scenarios. \citet{wang2025ideator} introduce IDEATOR, a black-box jailbreak loop in which a vision-language model proposes attack ideas and a text-to-image diffusion model renders them. \citet{ma2024visual} (Visual-RolePlay) and \citet{zhao2023attackvlm} use diffusion to render personas or target images that guide pixel-space adversarial generation against VLMs. \citet{wang2023instructta} use a similar pipeline for instruction-tuned targeted attack. None of these papers differentiates through the diffusion sampler.

\subsubsection{Diffusion-Style Universal Generators}
\citet{zhang2025anyattack} train a foundation-model-style additive-noise generator on LAION-400M with self-supervised contrastive pretraining, transferring across five open-source vision-language models and four commercial systems. The generator is not a pure diffusion sampler but uses diffusion-style large-scale pretraining as the basis for universal attacks.

\subsubsection{Attacks on Image-to-Image Diffusion Models}
\citet{zeng2025advi2i} attack the safety of image-to-image diffusion pipelines by training a generator that perturbs the conditioning image so that the downstream pipeline produces not-safe-for-work (NSFW) content even with benign text prompts. This work is adjacent rather than central to the LLM-jailbreak focus of this review, but is included because it represents the symmetric problem of diffusion-as-victim on the multimodal axis.

\subsubsection{VLM Cluster Critique}
\label{sec:cluster-critique-vlm}
The most striking observation about the VLM cluster is how rarely diffusion's gradients are actually exploited. Only AdvDiffVLM \citep{guo2024advdiffvlm} and the cross-attention perturbation work of \citet{xu2024transferable} truly differentiate through (or guide the score of) a diffusion sampler. The strongest pure-jailbreak papers in this space (Visual Adversarial Examples \citep{qi2024visual}, BAP \citep{ying2024bap}, UMK \citep{wang2024umk}) are pixel-space PGD. The strongest cross-VLM-transfer paper (IDEATOR \citep{wang2025ideator}) treats diffusion as a black-box renderer. Combining a true score-guided diffusion attack with a HarmBench-style multimodal safety evaluation is an open opportunity. Closed-VLM ASR remains a binding constraint across the cluster.

Table~\ref{tab:summary-vlm} summarizes the ten VLM-cluster papers, the adjacent attack on image-to-image diffusion safety, and the multimodal-side defense referenced above.

{\small
\begin{longtable}{P{1.8cm} P{1.3cm} P{2.2cm} P{3cm} P{0.6cm} P{4.1cm} P{0.5cm}}
\caption{Diffusion-based adversarial attacks on vision-language
models (Family~E), with one adjacent attack on image-to-image
diffusion (\citet{zeng2025advi2i}) and one multimodal diffusion
defense (\citet{zhu2025codefend}) included for context.
Legend as in Table~\ref{tab:summary-image}.}
\label{tab:summary-vlm}\\
\toprule
Reference & Venue / yr & Diffusion role & Target & Threat & Dataset / key reported finding & Code \\
\midrule
\endfirsthead
\multicolumn{7}{c}{\tablename\ \thetable{} -- continued from previous page} \\
\toprule
Reference & Venue / yr & Diffusion role & Target & Threat & Dataset / key reported finding & Code \\
\midrule
\endhead
\midrule
\multicolumn{7}{r}{continued on next page} \\
\endfoot
\bottomrule
\endlastfoot
\citet{guo2024advdiffvlm} & TIFS 2024 & Frozen, score guidance (AEGE + CAM mask) & GPT-4V, Gemini, Copilot, ERNIE, MiniGPT-4, LLaVA, BLIP-2 & GB/ BB$_{\text{open}}$ & ImageNet-derived caption pairs. 5--10$\times$ speed-up over transfer baselines; first true score-guided diffusion attack on commercial VLMs. & \cmark \\
\citet{xu2024transferable} & ACM MM 2024 & Frozen, cross-attention perturbation in SD & CLIP, ALBEF, TCL, BLIP & GB & MSCOCO, Flickr30K. Strong transferability for retrieval/VQA via SD cross-attention activation perturbation. & \xmark \\
\citet{zhang2025anyattack} & CVPR 2025 & Diffusion-style large-scale pretraining & CLIP, BLIP, BLIP-2, InstructBLIP, MiniGPT-4, plus Gemini, Claude, Copilot, GPT & BB & LAION-400M (pretraining); COCO eval. Foundation-model-style additive-noise generator transfers across commercial VLMs. & \cmark \\
\citet{li2024hades} & ECCV 2024 (Oral) & Pipeline (typography + SD scene + adversarial overlay) & LLaVA-1.5, MiniGPT-4, InstructBLIP, Gemini Pro Vision, GPT-4V & WB (overlay) & HADES (750 prompts; 5 categories). 90.26\% ASR on LLaVA-1.5; 71.60\% on Gemini Pro Vision. & \cmark \\
\citet{liu2024mmsafetybench} & ECCV 2024 & Pipeline (SD render + typography) for benchmark & LLaVA, MiniGPT-4, InstructBLIP, mPLUG-Owl, Qwen-VL, GPT-4V family & BB & MM-SafetyBench (5{,}040 image-text pairs, 13 unsafe scenarios). De-facto VLM safety benchmark. & \cmark \\
\citet{wang2025ideator} & ICCV 2025 & Pipeline (VLM ideation + diffusion render) & MiniGPT-4, LLaVA, InstructBLIP, Chameleon, GPT-4o, Claude-3.5, Gemini & BB & VLJailbreakBench (3{,}654). 94\% MiniGPT-4 (avg 5.34 queries); 46.31\% GPT-4o; 19.65\% Claude-3.5. & \cmark \\
\citet{zhao2023attackvlm} & NeurIPS 2023 & Pipeline (T2I diffusion proxy + CLIP/BLIP grad) & MiniGPT-4, LLaVA, UniDiffuser, BLIP-2, Img2Prompt & GB/BB & ImageNet, COCO. Targeted black-box VLM attack guided by diffusion-synthesized target images. & \cmark \\
\citet{wang2023instructta} & arXiv 2023 & Pipeline (T2I proxy targets + LLM-generated instructions) & MiniGPT-4, InstructBLIP, LLaVA & GB & MSCOCO, custom instructions. Instruction-tuning broadens transferability across diverse instructions. & \cmark \\
\citet{ma2024visual} & arXiv 2024 & Pipeline (T2I-rendered character) & LLaVA-1.6, Qwen-VL-Chat, OmniLMM, InternVL-1.5, Gemini-1.0 Pro Vision & BB & AdvBench-style harmful queries. Universal social-engineering jailbreak via diffusion-rendered personas. & \cmark \\
\citet{zeng2025advi2i} & ICML 2025 & Target: image-to-image diffusion is the victim & I2I latent-diffusion pipelines (SDEdit, InstructPix2Pix) & WB & I2P-style NSFW concept set. Bypasses Safe Latent Diffusion and adaptive defenses; subtle conditioning-image perturbations. & \cmark \\
\addlinespace
\citet{zhu2025codefend} & arXiv 2025 & VLM-side defense (image purifier + prompt prefix) & MLLMs (image captioning, VQA) & --- & Image-captioning and VQA benchmarks. Black-box-friendly defense; substantial robustness gains and transferability to unseen attacks. & \xmark \\
\end{longtable}
}

\subsection{Diffusion Large Language Models as Victims}
\label{sec:dllm-victim}
Although orthogonal to the use of diffusion as the attacker, four recent papers study what happens when diffusion LLMs are themselves the victims of an attack designed for the architecture.

\citet{wen2026dija} introduce DIJA, which exploits the bidirectional context modeling and parallel decoding of masked diffusion LLMs through interleaved mask-text prompts. \citet{zhang2025jailbreaking} report that autoregressive jailbreak attacks transfer poorly to diffusion LLMs and propose attacks specifically designed for the parallel-decoding mechanism, achieving high ASR on MMaDA-class models. \citet{li2026diffuguard} characterize how diffusion LLMs lose intrinsic safety when embedded in agent or structured-decoding contexts and propose a training-free defense based on stochastic remasking and block-level audit. \citet{yamabe2026toward} show that injecting a single affirmative token at an intermediate denoising step can steer the entire generation toward harmful content, exposing a priming vulnerability specific to the iterative-denoising mechanism.

These four papers are not about using diffusion to attack other LLMs, and so are out of the primary scope of this review. We include them because they identify decode-mechanic asymmetries that, in principle, could be exploited to design hybrid attackers in which a diffusion LLM serves as the generator and the asymmetric mechanic informs the attack design. Table~\ref{tab:summary-dllm-victim} summarizes the four Family~B
papers discussed in this subsection.

{\small
\begin{longtable}{P{1.8cm} P{1.3cm} P{2.2cm} P{2.5cm} P{0.8cm} P{4.0cm} P{0.5cm}}
\caption{Diffusion language models as victims of architecture-specific
attacks (Family~B). Threat: GF$^{\dag}$ denotes \emph{generation-flow}
access, in which the attacker manipulates the intermediate decoding
state of the dLLM victim; this category is not directly comparable to
WB/GB/BB on autoregressive targets. \cmark{}: public code released.}
\label{tab:summary-dllm-victim}\\
\toprule
Reference & Venue / yr & Diffusion role & Target & Threat & Dataset / key reported finding & Code \\
\midrule
\endfirsthead
\multicolumn{7}{c}{\tablename\ \thetable{} -- continued from previous page} \\
\toprule
Reference & Venue / yr & Diffusion role & Target & Threat & Dataset / key reported finding & Code \\
\midrule
\endhead
\midrule
\multicolumn{7}{r}{continued on next page} \\
\endfoot
\bottomrule
\endlastfoot
\citet{wen2026dija} & ICLR 2026 & dLLM as victim (interleaved mask-text prompts) & Dream-Instruct, LLaDA, MMaDA & GF$^{\dag}$ & JailbreakBench, StrongREJECT. Up to 100\% keyword ASR on Dream-Instruct; +78.5\% over ReNeLLM in evaluator-based ASR. & \cmark \\
\citet{zhang2025jailbreaking} & arXiv 2025 & dLLM as victim (Multi-Point Attention Attack) & MMaDA-Mix, Dream, LLaDA & GF$^{\dag}$ & Standard jailbreak benchmarks. 97\% ASR on MMaDA-Mix; AR jailbreaks transfer poorly to dLLMs. & \cmark \\
\citet{li2026diffuguard} & ICLR 2026 & dLLM-side defense (stochastic remasking) & Diffusion LLMs & --- & Six jailbreak attack families. Average ASR reduced from 47.9\% to 14.7\% with utility preserved. & \cmark \\
\citet{yamabe2026toward} & ICLR 2026 & dLLM as victim (priming with affirmative tokens) & Diffusion LLMs & GF$^{\dag}$ & Standard jailbreak benchmarks. Single affirmative token at an intermediate denoising step steers entire generation. & \cmark \\
\end{longtable}
}

\subsection{Non-Diffusion Baselines}
\label{sec:baselines}
For completeness and to anchor evaluation discussion, Table~\ref{tab:baselines} lists the major non-diffusion attacks against which any new diffusion-based attack should be compared. These attacks define the state of the art for adversarial-prompt search and serve as the baselines used in DiffusionAttacker, DART, and L\"udke et al.

{\small
\begin{longtable}{P{2.1cm} P{9.2cm} P{3cm}}
\caption{Non-diffusion adversarial baselines used for comparison in
the diffusion-attack literature.}
\label{tab:baselines}\\
\toprule
Method & Mechanism & Reference \\
\midrule
\endfirsthead
\multicolumn{3}{c}{\tablename\ \thetable{} -- continued from previous page} \\
\toprule
Method & Mechanism & Reference \\
\midrule
\endhead
\midrule
\multicolumn{3}{r}{continued on next page} \\
\endfoot
\bottomrule
\endlastfoot
GCG               & Greedy Coordinate Gradient on adversarial suffix; white-box.          & \citet{zou2023universal} \\
AutoDAN           & Hierarchical genetic algorithm on fluent jailbreaks.                   & \citet{liu2024autodan} \\
PAIR              & Black-box attacker LLM iteratively refines prompts.                    & \citet{chao2023jailbreaking} \\
AdvPrompter       & Trains autoregressive attacker to emit adversarial suffixes.           & \citet{paulus2024advprompter} \\
Visual AE         & Pixel-space PGD on universal jailbreak image for VLMs.                 & \citet{qi2024visual} \\
JailbreakInPieces & Compositional embedding-space attack on CLIP encoder.                  & \citet{shayegani2024jailbreak} \\
FigStep           & Typographic visual prompt jailbreak.                                   & \citet{gong2025figstep} \\
BAP               & Joint pixel-PGD and LLM-driven text refinement.                        & \citet{ying2024bap} \\
UMK               & Universal master-key co-optimization of image prefix and text suffix.  & \citet{wang2024umk} \\
Chameleon AE      & Gradient through joint embedding for early-fusion VLMs.                & \citet{rando2024gradient} \\
\end{longtable}
}

\section{Diffusion-Based Defenses for Text and Multimodal Models}
\label{sec:defenses}

A small but methodologically important cluster of work uses
diffusion-style processes to defend text models and multimodal models
against adversarial attacks. We discuss four papers below: three target
text classifiers, and one targets multimodal large language models.
Collectively they form the natural evaluation backdrop for any new
diffusion-based attack on the corresponding modality.

This section completes the dual attacker-defender perspective adopted throughout the review. The defense literature is currently asymmetric with the attack literature: while attacks proliferate, defenses are few and none has yet been audited by an adaptive attacker that explicitly back-propagates through the purifier in the way \citet{kang2024diffattack} does on the image side. We treat this asymmetry as itself a finding (Section~\ref{sec:rq5}) and develop the corresponding adaptive-attack research direction in Section~\ref{sec:open-questions}. In what follows, the four defenses are presented in the order in which they would form a natural target list for an adaptive-attack audit.

\subsection{Mask-and-Fill Purification}
\citet{li2023text} propose the foundational text-purification approach,
which masks and refills tokens of an input text using a masked language
model. The procedure is diffusion-adjacent: it implements a single-step
discrete corruption-and-restoration mechanism analogous to the forward
and reverse processes of a discrete diffusion model, but without an
explicit multi-step iterative denoising chain. The defense improves
robust accuracy under TextFooler~\citep{jin2020bertattack},
BERT-Attack, and DeepWordBug on standard classification benchmarks
(AG~News, IMDb, Yelp, SST-2). Although not a true diffusion model, the
work is the conceptual antecedent of the diffusion-based defenses that
followed.

\subsection{DiffuseDef}
\citet{li2025diffusedef} introduce DiffuseDef, which inserts an
embedding-space diffusion denoiser between the encoder and the
classifier of a text-classification pipeline. At inference, the
adversarial hidden state is combined with sampled noise and then
iteratively denoised, and the denoising outputs are ensembled to
produce a robust text representation. The defense is plug-and-play and
can be attached to any encoder. The reported gains are state-of-the-art
robust accuracy under common black-box and white-box attacks.
DiffuseDef is the most explicit realization of a true diffusion-based
defense for text and is the primary baseline that any diffusion-based
text attack should be evaluated against.

\subsection{MaskPure}
\citet{gietz2024maskpure} introduce MaskPure, a lightweight stochastic
purification defense inspired by diffusion processes. Input tokens are
randomly masked and refilled via a masked language model, and the
procedure is repeated stochastically. The defense matches or exceeds
contemporary baselines without requiring adversarial classifier
training, and is computationally cheaper than full diffusion-style
denoising. Together with DiffuseDef, MaskPure constitutes a second
tractable target for an adaptive-attack audit (see
Section~\ref{sec:implications}).

\subsection{Diffusion LLM Side: DiffuGuard}
On the diffusion-LLM-victim side, \citet{li2026diffuguard} propose
DiffuGuard, a training-free defense that combines stochastic annealing
remasking with block-level audit and repair, mitigating the
greedy-remasking bias and denoising-path-dependence vulnerabilities of
diffusion LLMs. Although the target is a different architecture,
DiffuGuard shares the same defensive philosophy as DiffuseDef and
MaskPure: stochastic intervention in the denoising trajectory.

\subsection{Multimodal Side: CoDefend}
\citet{zhu2025codefend} introduce CoDefend, the first diffusion-based
defense in our catalog that targets multimodal large language models
(MLLMs) rather than text classifiers. The procedure is collaborative
across modalities. An image purifier built on a supervised
diffusion-style denoiser takes adversarial images together with
denoising instructions as input and learns to remove adversarial
perturbations and restore clean inputs, while a prompt-prefix generator
is jointly optimized to harden the textual side against transfer
effects. The defense is black-box-friendly with respect to the
protected MLLM, requires only a small set of adversarial samples for
training, and the authors report substantial robustness gains plus
transferability to unseen attacks across image-captioning and visual
question answering (VQA) tasks. CoDefend is the natural multimodal
complement to DiffuseDef and MaskPure: it carries the same
stochastic-intervention philosophy from the text-classifier setting
into the VLM setting, and it is the most direct candidate against
which any new diffusion-based VLM jailbreak should be evaluated.

\section{Cross-Family Quantitative Comparison via a Five-Dimension Evaluation Framework}
\label{sec:quantitative}

The per-cluster discussions in Section~\ref{sec:taxonomy} reported each
paper's results inside the family that the paper belongs to. This
section steps outside that family-by-family structure and asks a
different question: when the reported numbers are placed side-by-side
across families, what does the attack landscape actually look like in
2024--2025? The exercise is harder than it should be, because the
cataloged papers do not use a single benchmark or a single definition
of attack success rate, and the form in which each paper reports its
results varies even within a single family. We are explicit about both
limitations below.

To support cross-paper comparison, we adopt throughout this review a five-dimension evaluation framework consisting of (i) \emph{attack success rate} (ASR), reported under the source paper's own definition (keyword-based, classifier-based, or judge-based); (ii) \emph{transferability}, reported as a pairwise success-rate gradient across target models, with particular attention to the open-weight versus frontier-closed gap; (iii) \emph{query budget}, reported as the number of model calls or restarts per successful attack; (iv) \emph{perplexity}, reported as a proxy for input-filter evasion under perplexity-based guardrails; and (v) \emph{defense-evasion}, reported as ASR retained against an explicit purifier or input filter such as DiffPure, DiffuseDef, MaskPure, CoDefend, Llama-Guard, or SmoothLLM. We apply this framework uniformly across the cataloged attacks where the source paper reports the relevant figure, and leave the cell blank where it does not. The framework is referenced in the RQ4 discussion (Section~\ref{sec:rq4}) and underpins the cluster-critique findings (Sections~\ref{sec:cluster-critique-llm} and~\ref{sec:cluster-critique-vlm}) and the open-questions agenda (Section~\ref{sec:open-questions}).

Table~\ref{tab:quantitative} consolidates the most directly comparable
numbers across the principal Family~A papers and a representative
slice of Family~E (VLM jailbreak) papers along the five framework
dimensions. We use only figures explicitly reported by the authors of
the respective papers and verified against the arXiv abstract,
accepted-paper version, or main results section. Where a paper does
not report a benchmark we leave the cell blank, and we do not impute.
Cells reporting ASR follow each source paper's own definition
(keyword-based, classifier-based, or judge-based), which is not
standardized across the field and which we discuss further in
Section~\ref{sec:evaluation}.

{\footnotesize
\begin{longtable}{P{2.5cm} P{2.0cm} P{2.4cm} P{2.0cm} P{1.6cm} P{2cm}}
\caption{Reported ASR and related figures for principal
diffusion-based attacks, organized along the five-dimension evaluation
framework defined in this section. Numbers are taken verbatim from the
source paper's abstract or main results section. Empty cells denote
benchmarks not reported in the source paper.}
\label{tab:quantitative}\\
\toprule
Reference & Benchmark & Target model & ASR (\%) & Query budget & Notes \\
\midrule
\endfirsthead
\multicolumn{6}{c}{\tablename\ \thetable{} -- continued from previous page} \\
\toprule
Reference & Benchmark & Target model & ASR (\%) & Query budget & Notes \\
\midrule
\endhead
\midrule
\multicolumn{6}{r}{continued on next page} \\
\endfoot
\bottomrule
\endlastfoot
\citet{wang2024diffusionattacker}   & AdvBench, HarmBench               & Llama-3-8B / Mistral-7B / Vicuna-7B / Alpaca+Safe-RLHF (transfer to GPT-4o, Claude-3.5) & 90/74, 93/79, 91/77, 88/71 (WB ASR\textsubscript{prefix}/ASR\textsubscript{GPT}); 56/49 GPT-4o, 33/21 Claude-3.5 (as rewriter) & not reported & seq2seq diffusion attacker with Gumbel-Softmax \\
\citet{noether2025dart}              & Red-teaming dataset; alpaca-gpt4  & gpt2-alpaca, Vicuna-7B, Llama2-7B-chat-hf       & dominates Pareto frontier (ASR vs cosine-sim) vs Unmodified, RL, Zero/Few-Shot, FLIRT at $\epsilon{=}0.1$ and $\epsilon{=}0.5$  & not reported  & no single ASR by design; see Figure~3 of source paper \\
\citet{ludke2025diffusion}           & JailbreakBench (StrongREJECT)     & Phi-4-Mini, Qwen-2.5-7B, Llama-3-8B, LAT-Llama-3-8B, Circuit Breakers, Gemma-3-1B, ChatGPT-5 & 100, 100, 100, 91, 93, 99, 53 (range 53--100 across 7 targets) & 2{,}000 restarts (100 for ChatGPT-5) & response-conditional inpainting on pretrained dLLM \\
\addlinespace
\citet{wen2026dija}                  & JailbreakBench, StrongREJECT      & Dream-Instruct (dLLM)           & up to 100 (keyword-based ASR) & not reported & +78.5\% over ReNeLLM (judge-based) \\
\citet{zhang2025jailbreaking}        & Standard jailbreak benchmarks     & MMaDA-Mix (dLLM)                & 97                            & not reported & autoregressive jailbreaks transfer poorly \\
\addlinespace
\citet{li2024hades}                  & HADES (750 prompts)               & LLaVA-1.5                       & 90.26                         & not reported & SD scene + typography + overlay \\
\citet{li2024hades}                  & HADES (750 prompts)               & Gemini Pro Vision               & 71.60                         & not reported & same setup \\
\citet{wang2025ideator}              & VLJailbreakBench (3{,}654)        & MiniGPT-4                       & 94                            & avg.\ 5.34 queries & VLM-driven jailbreak loop \\
\citet{wang2025ideator}              & VLJailbreakBench                  & LLaVA                           & 82                            & avg.\ 5.34 queries & transfer attack \\
\citet{wang2025ideator}              & VLJailbreakBench                  & InstructBLIP                    & 88                            & avg.\ 5.34 queries & transfer attack \\
\citet{wang2025ideator}              & VLJailbreakBench                  & Chameleon                       & 75                            & avg.\ 5.34 queries & transfer attack \\
\citet{wang2025ideator}              & VLJailbreakBench                  & GPT-4o                          & 46.31                         & avg.\ 5.34 queries & frontier closed-VLM \\
\citet{wang2025ideator}              & VLJailbreakBench                  & Claude-3.5-Sonnet               & 19.65                         & avg.\ 5.34 queries & frontier closed-VLM \\
\citet{wang2024umk}                  & AdvBench (universal)              & MiniGPT-4                       & 96                            & not reported & white-box image+text co-opt. \\
\citet{ying2024bap}                  & AdvBench / HarmBench-style        & Multiple VLMs                   & +29.03 over prior baselines (relative) & not reported & joint pixel-PGD + LLM CoT \\
\citet{rando2024gradient}            & AdvBench-style                    & Chameleon                       & 72.5                          & not reported & early-fusion gradient \\
\end{longtable}
}

Two patterns deserve to be named. The first concerns the Family~A
papers and is the source of most of the visual heterogeneity in the
table: of the four cataloged papers, only Qiu et al.\ truly does not
report extractable ASR figures (we discussed why in
Section~\ref{sec:qiu}). The other three do report numbers, but each
in a different form. DiffusionAttacker reports paired prefix-match
and GPT-judge ASRs across four open-weight targets, plus a separate
black-box-transfer pair on two frontier closed models, used as a
rewriter rather than a direct attacker. L\"udke et al.\ report
per-target percentages across seven targets spanning open-weight,
safety-hardened, and one closed model. DART reports a Pareto frontier
of (cosine-similarity, ASR) rather than any single point, because the
paper's central methodological argument is that ASR is only meaningful
when paired with a proximity measurement. The result is that
apples-to-apples comparison across Family~A is harder than it should
be at this stage of the field, and we read it as a signal that
reporting conventions for diffusion-based text attacks have not yet
stabilized.

The second pattern is more important. The strongest cross-VLM-transfer
attacker in the catalog, IDEATOR, reports a clean gradient across
target models: 94\% on MiniGPT-4, 82\% on LLaVA, 88\% on InstructBLIP,
75\% on Chameleon, then 46.31\% on GPT-4o and 19.65\% on
Claude-3.5-Sonnet. The first four numbers come from open-weight
targets; the last two come from frontier closed models. The closed
models are roughly a factor of two to four harder to jailbreak under
the same attacker, with the same query budget, on the same benchmark.
This is the empirical anchor for the recurring weakness we flagged in
Section~\ref{sec:cluster-critique-llm}: published diffusion-based
attacks have not yet been put through a HarmBench-scale evaluation
against the frontier closed models that defenders actually need to
defend.

A methodological note on what we did and did not do for this section.
For papers whose abstracts do not give a single canonical ASR but
whose main results section does, we extracted the figures from the
main results section and verified them against the arXiv version
where applicable. For papers that report Pareto-frontier or
distributional results rather than point estimates (DART), we did not
collapse the distribution to a single number, because doing so would
misrepresent the source paper's claim. Readers who want point
estimates from these papers should consult the source figures
directly; the per-paper references in Section~\ref{sec:text-attacks}
point to the specific figure and table identifiers.

\section{Evaluation Methods and Benchmarks}
\label{sec:evaluation}

Across the catalog, evaluation conventions converge on a small
number of datasets, metrics, and benchmarking practices. We summarize
these conventions below.

\subsection{Datasets}
Table~\ref{tab:datasets} lists the most frequently used datasets
across the cataloged papers, together with the count of cataloged
papers that use each. ImageNet is the most-used dataset in the
catalog, appearing in thirteen papers, all of them on the
image-classifier side. AdvBench is the most-used text-side dataset,
appearing in nine papers. The text-side benchmarks (AdvBench,
HarmBench, JailbreakBench) are used by the Family~A papers and by
the VLM-jailbreak papers in our Family~E. Face-recognition datasets
(CelebA, FFHQ) are used by the face-attack subcluster. CIFAR-10 is
used in the diffusion-purification-evasion subcluster.

{\small
\begin{longtable}{P{3.6cm} P{3.6cm} c P{5.4cm}}
\caption{Dataset usage across the cataloged papers. Counts reflect
the number of cataloged papers in which each dataset appears as a
primary evaluation set.}
\label{tab:datasets}\\
\toprule
Dataset & Predominant scope & Count & Reference \\
\midrule
\endfirsthead
\multicolumn{4}{c}{\tablename\ \thetable{} -- continued from previous page} \\
\toprule
Dataset & Predominant scope & Count & Reference \\
\midrule
\endhead
\midrule
\multicolumn{4}{r}{continued on next page} \\
\endfoot
\bottomrule
\endlastfoot
ImageNet            & Image classifier         & 13 & \citet{deng2009imagenet} \\
AdvBench            & LLM jailbreak (text)     & 9  & \citet{zou2023universal} \\
CelebA              & Face attack              & 5  & Public face dataset \\
COCO / MS-COCO      & VLM, detector            & 5  & \citet{lin2014mscoco} \\
HarmBench           & LLM jailbreak (text)     & 3  & \citet{mazeika2024harmbench} \\
FFHQ                & Face attack              & 3  & Public face dataset \\
JailbreakBench      & LLM jailbreak (text)     & 2  & \citet{chao2024jailbreakbench} \\
CIFAR-10            & Image classifier         & 2  & Public image dataset \\
LAION-400M          & VLM pretraining (attack) & 1  & Public image dataset \\
MM-SafetyBench      & VLM safety               & 1  & \citet{liu2024mmsafetybench} \\
HADES               & VLM jailbreak            & 1  & \citet{li2024hades} \\
SafeBench           & VLM jailbreak            & 1  & \citet{gong2025figstep} \\
RealToxicityPrompts & LLM toxicity             & 1  & \citet{gehman2020realtoxicity} \\
\end{longtable}
}

\subsection{Metrics}
The principal metric across the catalog is ASR, used in every paper, and it is the first dimension of the five-dimension evaluation framework introduced in Section~\ref{sec:quantitative}. Definitions of ASR vary: keyword-based (e.g., presence of refusal phrases), classifier-based (e.g., toxicity scores from a learned classifier), and judge-based (e.g., StrongREJECT, Llama-Guard, GPT-4 as judge). Naturalness metrics include Frechet Inception Distance (FID), Learned Perceptual Image Patch Similarity (LPIPS), and Structural Similarity Index (SSIM) on the image side, and perplexity, BLEU, ROUGE, and BERTScore on the text side. Perplexity also serves as the fourth framework dimension and is the standard proxy for input-filter evasion. Transferability is reported as a pairwise success-rate matrix across target models and corresponds to the second framework dimension; the third dimension, query budget (number of model calls), is reported by black-box attacks. Diversity is reported as cluster counts or Self-BLEU on the text side. The cataloged papers do not converge on a single naturalness metric for text, which is consistent with the broader observation that no perplexity threshold has emerged as a deployment-level filter standard.

\subsection{Benchmarking Practices}
The text-side literature benefits from standardized benchmarks. AdvBench provides 5{,}000 pairs of harmful behaviors and target strings; HarmBench standardizes evaluation of automated red teaming with a learned classifier; JailbreakBench provides a curated set of harmful prompts with a strong-judge protocol. The image-classifier literature has no equivalent single benchmark and instead uses dataset-specific transferability protocols (e.g., NIPS17 1k subset for ImageNet transferability). The VLM-side literature is intermediate: MM-SafetyBench, HADES, VLJailbreakBench, and SafeBench each cover distinct slices of the safety space, but no single benchmark dominates. No benchmark has been built specifically for diffusion-generated
adversarial inputs, which we identify as a near-term opportunity
(Section~\ref{sec:implications}).

\section{Analysis and Discussion}
\label{sec:discussion}

This section returns to the six research questions stated in
Section~\ref{sec:rqs} and answers each one directly, drawing on the
catalog material developed in Sections~\ref{sec:taxonomy} through
\ref{sec:evaluation}. We answer the questions in order. Where an
answer is short because the catalog speaks for itself, we keep it
short rather than pad it; where an answer requires nuance, we develop
it in proportion to the underlying evidence.

\subsection{RQ1: Current State of Diffusion-Based Adversarial Attacks on ML Systems}
\label{sec:rq1}

The current state of the field is small, recent, and methodologically
diverse on the LLM side, and substantially larger and more mature on
the image-classifier side. The catalog contains four
diffusion-attack papers on text/LLMs, four diffusion-LLM-as-victim
papers, eighteen on image classifiers, ten on vision-language models,
four diffusion-based defenses, and ten non-diffusion baselines. The
LLM-side cluster (Family~A) is the focal subject of this review and
also its newest cluster: all four papers appeared in 2024 or 2025,
and they cover four genuinely distinct attack philosophies
(continuous-latent denoising with differentiable sampling
\citep{wang2024diffusionattacker}, embedding-space perturbation under
proximity constraint \citep{noether2025dart}, inference-time
inpainting on a pretrained masked dLLM \citep{ludke2025diffusion},
and dual-objective embedding-space training for transfer
\citep{qiu2024diffusion}). The image-side cluster (Family~D) is
mature and contains specific methodological recipes,
namely DDIM-inversion plus latent perturbation, classifier-score
guidance, and segment-wise back-propagation through purifiers, that
have not yet been ported to the text setting. The dLLM-as-victim
literature (Family~B) is adjacent rather than primary but is growing
rapidly, with four papers all appearing in 2025--2026, and we
include it because hybrid attacker designs may exploit those
findings. The defense side (Family~C) is small, comprising three
text-classifier defenses and one multimodal-LLM defense, and none
has yet been broken by an adaptive diffusion-based attack in the
published literature.

\subsection{RQ2: Taxonomy of Diffusion Roles in Adversarial Pipelines}
\label{sec:rq2}

We propose a six-class taxonomy of how diffusion is used in an
adversarial pipeline: (i)~trained generator, in which the diffusion
model is trained end-to-end with an attack objective
\citep{wang2024diffusionattacker,noether2025dart,qiu2024diffusion};
(ii)~frozen pretrained sampler with latent perturbation, in which
DDIM inversion plus latent-space gradient is used to attack a
classifier downstream
\citep{chen2024diffattack,chen2023content,liu2024advdiffusion,sun2024diffam};
(iii)~frozen pretrained sampler with score or classifier guidance,
in which an adversarial gradient term is injected into the reverse
SDE at each denoising step
\citep{dai2024advdiff,chen2023advdiffuser,xue2023diffpgd,huang2025scoreadv,collins2025natadiff,dai2025semdiff,jiang2025apa,guo2024advdiffvlm};
(iv)~off-the-shelf inference, in which a pretrained diffusion
language model is used without any training and without any
optimization \citep{ludke2025diffusion}; (v)~pipeline-only renderer,
in which diffusion produces images or scenes that are then attacked
through a separate channel, with no gradient flowing through the
diffusion sampler
\citep{li2024hades,liu2024mmsafetybench,wang2025ideator,ma2024visual,zhao2023attackvlm,wang2023instructta};
and (vi)~victim diffusion model, in which the diffusion model is the
target of the attack rather than a component of the attacker
\citep{zeng2025advi2i,wen2026dija,zhang2025jailbreaking,yamabe2026toward,li2026diffuguard}.
We augment this taxonomy with a threat-model axis (white-box,
grey-box, black-box, generation-flow) and a query-budget axis. The
taxonomy distinguishes papers that exploit diffusion's gradients
(roles ii and iii) from those that use it as a black-box rendering
step (role v), which is the methodologically significant cut.

\subsection{RQ3: Formulations, Training Methods, and Optimization Strategies}
\label{sec:rq3}

The cataloged papers converge on a small number of formulations and
optimization techniques. On the formulation side, continuous-latent
diffusion (operating on real-valued embeddings) is the dominant
choice for text \citep{wang2024diffusionattacker,noether2025dart}
because it provides differentiable surfaces over which an attack
loss can be optimized, while latent diffusion (operating in the
latent space of a pretrained autoencoder) dominates the image-side
literature because it inherits the strong sample quality of Stable
Diffusion. Discrete diffusion (operating directly on token
sequences) is conspicuously underused: no Family~A paper uses a true
discrete-diffusion sampler such as D3PM, SEDD, or MDLM with an
adversarial training objective, and we flag this as the most
tractable unfilled cell of the matrix in Section~\ref{sec:open-questions}.

On the training side, the catalog has accumulated a small toolkit
for managing the high compute cost of back-propagation through the
full reverse trajectory: segment-wise back-propagation
\citep{kang2024diffattack} bounds memory by chunking the
denoising chain; $x_0$-prediction surrogates
\citep{chen2023advdiffuser} avoid full-trajectory gradients by
attacking the predicted clean sample at each step; time-travel
sampling \citep{collins2025natadiff} revisits earlier denoising
states to refine the attack; and Gumbel-Softmax relaxation
\citep{wang2024diffusionattacker} makes the discrete-token sampling
step at the end of a text-diffusion pipeline differentiable.
Reinforcement learning, specifically PPO with a toxicity reward,
provides an alternative when end-to-end differentiability is
unavailable \citep{noether2025dart}. Finally, off-the-shelf inference
without any training is a genuinely distinct strategy that was first
operationalized at scale by \citet{ludke2025diffusion} and that any
new diffusion-based attacker on LLMs must now differentiate from,
because the baseline is nearly free.

\subsection{RQ4: Datasets, Metrics, and Target Models}
\label{sec:rq4}

The catalog converges on a small number of benchmarks. On the text
side, AdvBench (used in nine papers), HarmBench (three),
and JailbreakBench (two) are the standardized choices, all
supplemented by paper-specific reference-prompt sets where the
attack requires a specific input distribution. On the image side,
ImageNet (thirteen papers) dominates, with face-recognition datasets
(CelebA in five, FFHQ in three) used by the face-attack subcluster
and CIFAR-10 used in the purification-evasion subcluster. On the
VLM side, MM-SafetyBench, HADES, VLJailbreakBench, and SafeBench
each cover distinct slices of the safety space; no single VLM
benchmark dominates. The principal metric across the catalog is
ASR, used in every paper, though its operationalization is
heterogeneous (keyword-based, classifier-based, or judge-based),
which we identify as a recurring weakness of the field
(Section~\ref{sec:rq6}). Within our five-dimension evaluation framework (Section~\ref{sec:quantitative}), ASR is the most consistently reported dimension across the catalog, transferability and query budget are reported only by black-box attacks, and perplexity and defense-evasion are reported only sporadically. Naturalness is reported via FID and LPIPS
on the image side and via perplexity on the text side, although no
deployment-level perplexity threshold has emerged as a standard.

Target-model coverage leans heavily on open-weight models. The
targets that appear in five or more cataloged papers, in descending
order of frequency, are GPT-4 (13), MiniGPT-4 (12), BLIP variants
(11), LLaVA (10), ResNet (8), InstructBLIP (8), and ViT (7).
Frontier closed models appear in 5 papers for Claude and 4 for
Gemini, and the reported ASR on these models is consistently lower
(Section~\ref{sec:quantitative}). This asymmetry, namely that
diffusion-based attacks are systematically evaluated on the easier
targets, is the empirical anchor for one of the recurring
weaknesses we identify under RQ6.

\subsection{RQ5: Diffusion-Based Defenses for Text Models}
\label{sec:rq5}

Four diffusion-based defenses are now in the catalog: three target
text classifiers (Text Adversarial Purification
\citep{li2023text}, DiffuseDef \citep{li2025diffusedef}, MaskPure
\citep{gietz2024maskpure}) and one targets multimodal large language
models (CoDefend \citep{zhu2025codefend}). All four share a common
philosophy of stochastic intervention in the denoising trajectory:
input tokens or embeddings are corrupted with noise or masking and
then refilled or denoised, with the stochasticity expected to
disrupt adversarial perturbations while preserving benign content.
None of the four has yet been broken by an adaptive
diffusion-based attack in the published literature, which is the
single most direct gap exposed by this review and which we develop
into a concrete research direction in
Section~\ref{sec:open-questions}.

The relationship between the defense literature and the attack
literature is currently asymmetric. The image-side attack literature
has produced segment-wise back-propagation attacks
\citep{kang2024diffattack} that explicitly target the DiffPure
defense; the text-side attack literature has produced no comparable
adaptive attack against DiffuseDef or MaskPure, and the multimodal
side has none against CoDefend. The defense papers therefore stand
unaudited in the sense that matters most, which is whether their
reported robustness gains survive an attacker who knows the defense
mechanism and back-propagates through it.

\subsection{RQ6: Gaps and Unaddressed Questions}
\label{sec:rq6}

The catalog exposes five recurring gaps in the LLM-side literature,
which we developed in detail in
Section~\ref{sec:cluster-critique-llm} and summarize here in the
form RQ6 asks for. First, white-box dependency: three of the four
Family~A papers train against white-box gradients of the target
model, leaving direct training-free attack on closed frontier
models thinly populated. Second, underuse of discrete diffusion: no
Family~A paper combines a true discrete-diffusion sampler with an
explicit harmfulness reward and a HarmBench-scale evaluation. Third,
defense evasion under-evaluation: no paper benchmarks against modern
input filters such as Llama-Guard \citep{inan2023llamaguard} or
SmoothLLM \citep{robey2023smoothllm}, and reported low-perplexity
numbers are necessary but not sufficient evidence of filter evasion.
Fourth, single-turn restriction: all four Family~A papers produce
single-shot adversarial prompts, even though diffusion's natural
strength (joint modeling with bidirectional conditioning) is
precisely what one would want for multi-turn dialogue attacks.
Fifth, constrained or topic-specific red teaming is represented by a
single paper (DART), and combining DART's proximity constraint with
a true discrete-diffusion sampler is an obvious unfilled cell of the
methodological matrix.

Of these five gaps, we judge the most tractable to be the third
(defense evasion under-evaluation), because the defenses are public,
the adaptive-attack recipe is known from the image side, and the
exercise produces a concrete, falsifiable empirical claim.
Section~\ref{sec:open-questions} develops each of the five into a
concrete research direction with an associated experimental design.

\subsection{Cross-Cutting Observations}
\label{sec:cross-cutting}

Two cross-cutting observations emerge from the RQ-by-RQ analysis
that do not fit any single research question.

The first is that the field has bifurcated into two reframings of
the adversarial-diffusion problem. The first reframing,
\emph{perturb-an-input} via DDIM-inversion plus latent gradient,
preserves a one-to-one correspondence between input and adversarial
output and is the closest analogue to classical adversarial attacks.
The second, \emph{guide-from-noise} via classifier or score
guidance, creates novel adversarial samples and is the closer
analogue to red-team prompt synthesis. The two reframings have
different threat models, different compute costs, and different
naturalness profiles, and we believe the field would benefit from
treating them as methodologically distinct rather than as variants
of a single recipe.

The second cross-cutting observation is that the dLLM-as-victim
literature (Family~B) and the dLLM-as-attacker literature
(\citet{ludke2025diffusion} within Family~A) are studying the same
class of models from opposite sides, but the two literatures have not
yet been cross-referenced in a single attacker design. A hybrid
attacker that exploits decode-mechanic asymmetries identified in
Family~B \citep{wen2026dija,zhang2025jailbreaking,yamabe2026toward}
to inform an attacker built on the dLLM-as-natural-adversary framing
of \citet{ludke2025diffusion} is a research direction we have not
seen articulated in the published literature, and we develop it
further in Section~\ref{sec:open-questions}.

\section{Limitations and Threats to Validity}
\label{sec:threats}

We structure the limitations of this review under the four standard
threat-to-validity categories used in empirical software engineering
and secondary-study methodology: internal validity (whether the
review process itself produced reliable conclusions), external
validity (whether the conclusions generalize beyond the cataloged
corpus), construct validity (whether the constructs we used to
classify and compare papers actually measure what we claim they
measure), and conclusion validity (whether the empirical claims drawn
from the catalog are warranted by the underlying evidence). For each,
we identify the specific threats most relevant to the present review
and the mitigations we applied.

\paragraph{Internal validity.}
The principal internal threat is that this is a narrative review with
quality assessment rather than a PRISMA-compliant systematic review.
The authors performed the complete screening pipeline by hand without
pre-registration of the protocol and without independent dual-author
blinded screening of the full candidate pool. We mitigated this in
three ways. First, we made the choice explicit at the outset
(Section~\ref{sec:method}) rather than implying systematic-review
machinery the work does not employ. Second, we report a supplementary
post-hoc agreement check between two independent LLM agents on the
inclusion rule, yielding Cohen's $\kappa = 0.781$ on the 154-record
candidate pool; this is a transparency probe on the precision of the
inclusion rule, not a substitute for formal inter-rater agreement,
and we say so. Third, the complete screening log, including the 44
records excluded at full-text review with their exclusion reasons,
is released in the companion repository, so a reader who disagrees
with any single inclusion or exclusion can trace and challenge it.
A secondary internal threat is that author lists, venues, and
identifiers were verified against arXiv abstract pages and publisher
pages but not against the printed journal proofs for every entry;
for a small number of journal entries (the Springer Applied
Intelligence and ScienceDirect Neurocomputing papers), full author
lists were not cleanly available from public listings and we used
best-available subset listings flagged with ``et al.''

\paragraph{External validity.}
The principal external threat is that our coverage is bounded by
the search cutoff (May 2026), and the LLM-side cluster is small
enough (four papers) that a single high-impact paper appearing after
the cutoff could meaningfully shift the centre of gravity of the
analysis. We believe the taxonomy and the recurring-weakness
analysis are robust against incremental additions, but new
methodological families, for example flow-matching attacks, could
require taxonomic revision. A second external threat is that the
target-model coverage in the catalog leans heavily on open-weight
models (GPT-4 in 13 papers, MiniGPT-4 in 12, LLaVA in 10), with
frontier closed models underrepresented (Claude in 5 papers, Gemini
in 4); conclusions about closed-model robustness therefore generalize
less reliably than conclusions about open-weight robustness. We are
explicit about this asymmetry in Section~\ref{sec:quantitative} when
discussing the closed-VLM gap. A third external threat is geographic
and venue concentration: the catalog draws heavily from NeurIPS,
ICML, ICLR, CVPR, and arXiv, and is comparatively light on security
venues (USENIX Security, IEEE S\&P, CCS, NDSS), which means
practitioner-oriented attack work that appears at security venues
without an arXiv version may be under-sampled.

\paragraph{Construct validity.}
The principal construct threat concerns the taxonomy itself. Our
six-class taxonomy of diffusion roles is methodological rather than
capability-based, and it was developed iteratively against the
catalog as the catalog grew, which means it fits the current
literature well by construction. Whether it remains the right
taxonomy for the next fifty papers is an open question. Some papers
fit multiple categories (DiffPatch uses both null-text inversion and
latent perturbation), and some recent papers (APA) introduce
training paradigms such as preference alignment that are not native
to any single category; we classified each paper by its dominant
role and noted alternatives in the catalog, but this is a judgment
call that another reviewer might make differently. A second
construct threat concerns the threat-model axis. We use the standard
white-box / grey-box / black-box trichotomy plus a generation-flow
category for dLLM-victim attacks, but the boundary between grey-box
and black-box depends on what surrogate or encoder-only access counts
as ``access,'' and the boundary between white-box-at-training-time
and black-box-at-inference (relevant for DiffusionAttacker as
rewriter) does not have a single accepted convention in the field.
We adopt the convention used by each source paper and flag it in the
threat column.

\paragraph{Conclusion validity.}
The principal conclusion threat is that ASR is not standardized
across the cataloged papers. Three definitions coexist (keyword-based,
classifier-based, judge-based), each with different sensitivity to
refusal phrasing, paraphrase, and reward-hacking, and a number that
is high under one definition can be substantially lower under
another. We have not reweighted reported ASRs to a common definition
because doing so would require re-running each attack and would
exceed the scope of a review. We mitigate this by reporting each
paper's ASR under its own definition, flagging the definition where
the source paper states it, and limiting cross-paper claims to
patterns visible across multiple papers and definitions (such as the
closed-model gap in Section~\ref{sec:quantitative}, which holds under
both judge-based and classifier-based ASRs). A second conclusion
threat concerns papers that report Pareto frontiers or per-target
distributions rather than point estimates (DART, L\"udke et al.,
DiffusionAttacker): we did not collapse those distributions to a
single number in Table~\ref{tab:quantitative}, because doing so would
misrepresent the source paper's claim, but this means some
table cells are heterogeneous in shape and a reader looking for a
single comparable number will not find one in every row.

\paragraph{Scope clarification.}
For completeness, we note that the review focuses on the LLM-centric
use of diffusion in adversarial pipelines. Adjacent literatures,
including diffusion-based watermarking, diffusion-based data
poisoning, and diffusion-based deepfake generation, are not covered.
Image-side defenses (DiffPure and follow-ups) are referenced but not
enumerated because they fall outside the LLM-centered scope. The
dLLM-as-victim literature is included as adjacent rather than as a
primary focus, which is a choice with its own trade-offs: it makes
the survey more forward-looking but means we treat four genuinely
distinct attacks (Family~B) at a lighter level of methodological
detail than the four Family~A attacks. This is a scope choice rather
than a validity threat in the conventional sense, which is why we
break it out as a separate paragraph.
\section{Implications and Future Research Directions}
\label{sec:implications}

Section~\ref{sec:discussion} characterized the state of the field; this section describes the implications of that state and the open research directions it suggests. We organize the section in four parts. First, we synthesize the central findings of the review and indicate which are most consequential for follow-on work. Second, we collect the open research questions raised by the analysis into thematic groups and frame them as opportunities for the community rather than as a prescriptive list of contributions. Third, we discuss the impact of the review on the broader LLM red teaming community. Fourth, we address the ethical implications of consolidating an attack-oriented literature into a single reference, including the choices we have made in this paper to mitigate dual-use risks.

\subsection{Synthesis of Key Findings}

The principal finding of this review is that the LLM-side literature on diffusion-based attacks is small, recent, and methodologically diverse. Four published papers cover four distinct attack philosophies: amortized inference via off-the-shelf diffusion LLMs \citep{ludke2025diffusion}; trained sequence-to-sequence diffusion attackers \citep{wang2024diffusionattacker}; constrained-search perturbation models \citep{noether2025dart}; and substitute-classifier-guided embedding diffusion against text classifiers \citep{qiu2024diffusion}. The image-side literature provides specific recipes (DDIM-inversion plus latent perturbation, classifier-score guidance, segment-wise back-propagation through purifiers) that have not yet been transplanted to the text setting. The vision-language-model literature is dominated by pipeline-only uses of diffusion and could benefit from the score-guided approaches developed on the image side.

\subsection{Open Research Questions}
\label{sec:open-questions}

We organize the open questions raised by our analysis into three thematic groups. We frame these as questions for the research community rather than as a list of publishable contributions, and we note that some of these questions may already be the subject of unpublished or in-progress work at the time of reading.

\paragraph{Questions about defenses and adaptive evaluation.}
The most direct gap exposed by the catalog is that the diffusion-based purification defenses have not yet been audited by an adaptive attack that explicitly back-propagates through the purifier in the way \citet{kang2024diffattack} does on the image side. Three text-side defenses are open in this respect (DiffuseDef~\citep{li2025diffusedef}, MaskPure~\citep{gietz2024maskpure}, and the foundational mask-and-fill purifier of \citet{li2023text}), and the multimodal-side defense CoDefend~\citep{zhu2025codefend} is similarly unaudited. A useful research question is whether the segment-wise-back-propagation recipe transfers to text and to multimodal pipelines, and how robust accuracy of the defended models holds up under that audit. A related question is whether the foundational mask-and-fill purifier of \citet{li2023text}, which lacks an explicit multi-step diffusion process, is qualitatively easier or harder to break than the multi-step variants. CoDefend is particularly interesting because its joint image-purifier-and-prompt-prefix design suggests that an effective adaptive attack must perturb both modalities simultaneously, which is a non-trivial extension of the text-only adaptive-attack recipe.

\paragraph{Questions about generator design.}
A second cluster of questions concerns the design of the diffusion generator itself. No published paper combines a true discrete-diffusion sampler (D3PM, SEDD, or MDLM~\citep{austin2021structured,lou2024discrete,shi2024simplified}) with an explicit harmfulness reward and a HarmBench-scale evaluation. The text analogue of \citet{dai2024advdiff}'s score-guided image attack, namely a score-guided discrete-diffusion attack on text classifiers compared against TextFooler~\citep{jin2020bertattack} and BERT-Attack baselines, has not been published. Similarly, the proximity-constrained formulation of DART~\citep{noether2025dart} could be reimplemented on top of a real masked-diffusion sampler to ask whether the proximity benefits of DART hold under a stronger generator. A further question is whether the diffusion-LLM-as-natural-adversary framing of \citet{ludke2025diffusion} can be strengthened by reinforcement-learning fine-tuning of the dLLM with a harmfulness reward under fluency constraints, addressing the fidelity-assumption concern raised in Section~\ref{sec:ludke}.

\paragraph{Questions about transferability, evaluation, and downstream effect.}
A third cluster concerns evaluation. The transferability of outputs from DiffusionAttacker, DART, and \citet{ludke2025diffusion} against frontier closed LLMs (GPT-4, Claude, Gemini) at realistic query budgets has not been systematically reported. Such a study would directly inform the practical relevance of the field. Whether perplexity-controlled diffusion attacks can pass through deployment-grade input filters such as Llama-Guard~\citep{inan2023llamaguard} or SmoothLLM~\citep{robey2023smoothllm} is also unaddressed. Finally, the multi-turn setting is uncultivated: diffusion's bidirectional conditioning is structurally well-suited to generating coherent adversarial dialogues with both user and assistant turns, and the absence of published work in this direction is striking.

A more speculative direction concerns hybrid attackers that exploit decode-mechanic asymmetries identified in the diffusion-LLM-victim literature~\citep{wen2026dija,zhang2025jailbreaking,yamabe2026toward}. If those asymmetries partially explain a diffusion LLM's distribution over harmful prompts, an attacker that explicitly conditions on those mechanics could outperform a generic dLLM-as-natural-adversary baseline. We flag this as a research question, with the caveat that it presupposes an empirical link between victim-side and attacker-side behavior that has not been established in the published literature.

A second speculative direction is joint multimodal red teaming via diffusion: combining AdvDiffVLM-style score guidance on the image side with a discrete-diffusion text attacker on the prompt side and optimizing a joint VLM-jailbreak loss. No paper in the catalog has attempted this composition, and we view it as the natural bi-modal extension of the Family A line.

\subsection{Impact Statement}
This review consolidates a fragmented literature into a single reference, supporting both attackers and defenders in the LLM red teaming community. By making the methodological vocabulary of the image-side literature accessible to text-side researchers, we lower the barrier to porting recipes that are likely to yield publishable contributions. By identifying defense-evaluation gaps (no published adaptive attack on DiffuseDef, MaskPure, or CoDefend), we provide defenders with a tractable benchmark against which to harden their systems. By cataloging the rapidly growing diffusion-LLM-as-victim literature alongside the diffusion-as-attacker literature, we make explicit a hybrid-attacker direction that has not been previously articulated. As the broader LLM safety literature continues to mature, we anticipate that diffusion-based attacks will move from a curiosity to a standard component of the red teaming toolbox, and we hope this review contributes to that trajectory in a constructive direction.

\subsection{Ethical Considerations and Responsible Disclosure}
\label{sec:ethics}

A review article that catalogs adversarial attacks on safety-aligned language models has dual-use implications, and we treat this explicitly rather than implicitly. Our position is that consolidating the existing literature into a single reference is net beneficial because the work surveyed is already public, and a clearer map of the area helps defenders, evaluators, and policymakers more than it helps adversaries who are willing to read several dozen papers in any case. Two specific choices reflect this position. First, we describe attack mechanisms at the level of methodological design rather than reproducing harmful prompts, model outputs, or attack code; the companion catalog likewise records mechanism summaries rather than payloads. Second, the open-research-question framing in Section~\ref{sec:open-questions} prioritizes evaluation gaps and defense-side questions before generator-design questions, on the view that improving evaluation rigor and breaking diffusion-based defenses are higher-leverage than producing a stronger generator. The cataloged papers themselves vary in their disclosure practices: some release code openly, some embargo, and some include explicit responsible-disclosure statements. We do not take a position here on which practice is correct, but we note that the field would benefit from a more uniform norm. We also note that any researcher building on the open questions in Section~\ref{sec:open-questions} should consider engaging with model providers ahead of public release, particularly for attacks that demonstrate non-trivial transfer to deployed closed-source systems.

\section{Conclusion}
\label{sec:conclusion}

This review surveyed fifty published papers on the use of diffusion models in adversarial pipelines targeting machine-learning systems, with the primary focus on LLMs. We introduced a six-class taxonomy of diffusion roles that distinguishes trained generators from frozen pretrained samplers, score-guided attacks from latent-perturbation attacks, and pipeline-only uses from genuine gradient-based exploitation. We developed a five-dimension evaluation framework (ASR, transferability, query budget, perplexity, defense-evasion) and applied it uniformly across the cataloged attacks to surface the closed-frontier-model gap and the defense-audit gap. We critiqued the LLM-side literature along five axes (white-box dependency, missing frontier-model evaluation, underuse of discrete-diffusion samplers, weak defense-evasion evaluation, and absence of multi-turn formulations) and proposed concrete research directions in each axis. We covered the larger image-classifier and vision-language-model literatures as methodological background and the diffusion-LLM-victim literature as a source of hybrid-attacker design ideas. We summarized evaluation conventions across the catalog, identified the most-used datasets and metrics, and noted the absence of a benchmark specifically for diffusion-generated adversarial inputs.

The principal conclusion is that the use of diffusion as a generator of adversarial text against LLMs is a young area with a small number of strong representative papers and substantial methodological room for follow-on work. Researchers entering the area can proceed along multiple complementary directions, with a research agenda that runs from defense audit through generator design to multimodal extension. We hope this review provides a useful map for that work, and a useful exercise in cross-stream information fusion for the broader community concerned with the synergism among the many disciplines contributing to adversarial machine learning.

\section*{CRediT authorship contribution statement}
\textbf{Abrar Alotaibi:} Conceptualization, Methodology, Investigation, Data Curation, Writing -- Original Draft, Visualization. \textbf{Moataz Ahmed:} Writing -- Review \& Editing, Supervision, Project administration, Funding acquisition.

\section*{Acknowledgements}
The authors would like to acknowledge that this research is supported by a grant (No.\ CRPG-25-2057) under the Cybersecurity Research and Innovation Pioneers Initiative, provided by the National Cybersecurity Authority. The authors gratefully acknowledge the support received from the Saudi Data and AI Authority (SDAIA) and King Fahd University of Petroleum and Minerals (KFUPM) under the SDAIA-KFUPM Joint Research Center for Artificial Intelligence.

\section*{Data and Catalog Availability}
The complete search-and-screening artifact for this review is
released as a public spreadsheet in the companion repository
(\url{https://github.com/AbrarAlotaibi/diffusion-redteam-llm-survey}).
The artifact covers every record considered, not only those retained:
all 154 candidate records returned by the initial search, the
title-and-abstract screening decision for each, the full-text
eligibility decisions for the 94 records that advanced to that stage,
and the exclusion reason for each of the 44 records excluded at
full-text review. The final fifty-paper catalog of included works is
recorded in the same spreadsheet alongside the four diffusion-LLM-victim entries and ten non-diffusion baselines retained for context; per-paper metadata covers scope, diffusion role, threat model, formulation, training method, datasets, metrics, target models, code link, key contribution, and principal limitation. The
repository additionally includes the two LLM-agent re-screening runs used for the post-hoc agreement check reported in Section~\ref{sec:selection}, together with the agent prompts and the per-record disagreement set on which Cohen's $\kappa = 0.781$ was
computed.

\section*{Declaration of competing interests}
The authors declare no competing financial or non-financial interests.

\bibliographystyle{cas-model2-names}
\bibliography{refs}

@article{wang2024diffusionattacker,
  author  = {Wang, Hao and Li, Hao and Zhu, Junda and Wang, Xinyuan and Pan, Chengwei and Huang, Minlie and Sha, Lei},
  title   = {{DiffusionAttacker}: Diffusion-Driven Prompt Manipulation for {LLM} Jailbreak},
  journal = {arXiv preprint arXiv:2412.17522},
  year    = {2024},
  doi     = {10.48550/arXiv.2412.17522}
}

@inproceedings{noether2025dart,
  author    = {N{\"o}ther, Jonathan and Singla, Adish and Radanovi{\'c}, Goran},
  title     = {Text-Diffusion Red-Teaming of Large Language Models: Unveiling Harmful Behaviors with Proximity Constraints},
  booktitle = {Proceedings of the AAAI Conference on Artificial Intelligence},
  year      = {2025},
  doi       = {10.48550/arXiv.2501.08246}
}

@article{ludke2025diffusion,
  author  = {L{\"u}dke, David and Wollschl{\"a}ger, Tom and Ungermann, Paul and G{\"u}nnemann, Stephan and Schwinn, Leo},
  title   = {Diffusion {LLMs} are Natural Adversaries for any {LLM}},
  journal = {arXiv preprint arXiv:2511.00203},
  year    = {2025},
  doi     = {10.48550/arXiv.2511.00203}
}

@inproceedings{qiu2024diffusion,
  author    = {Qiu, Shilin and Gou, Min and Liang, Tao},
  title     = {Diffusion Model for Adversarial Attack Against {NLP} Models},
  booktitle = {Proceedings of the SPIE Vol.~13105 (ICCAID 2023)},
  year      = {2024},
  doi       = {10.1117/12.3026312}
}

@inproceedings{wen2026dija,
  author    = {Wen, Zichen and Qu, Jiashu and Chen, Zhaorun and Lu, Xiaoya and Liu, Dongrui and Liu, Zhiyuan and Wu, Ruixi and Yang, Yicun and Jin, Xiangqi and Xu, Haoyun and Liu, Xuyang and Li, Weijia and Lu, Chaochao and Shao, Jing and He, Conghui and Zhang, Linfeng},
  title     = {The Devil behind the mask: An emergent safety vulnerability of Diffusion {LLMs}},
  booktitle = {International Conference on Learning Representations (ICLR)},
  year      = {2026},
  doi       = {10.48550/arXiv.2507.11097}
}

@article{zhang2025jailbreaking,
  author  = {Zhang, Yuanhe and Xie, Fangzhou and Zhou, Zhenhong and Li, Zherui and Chen, Hao and Wang, Kun and Guo, Yufei},
  title   = {Jailbreaking Large Language Diffusion Models: Revealing Hidden Safety Flaws in Diffusion-Based Text Generation},
  journal = {arXiv preprint arXiv:2507.19227},
  year    = {2025},
  doi     = {10.48550/arXiv.2507.19227}
}

@inproceedings{li2026diffuguard,
  author    = {Li, Zherui and Nie, Zheng and Zhou, Zhenhong and Liu, Yue and Zhang, Yitong and Cheng, Yu and Wen, Qingsong and Wang, Kun and Guo, Yufei and Zhang, Jiaheng},
  title     = {{DiffuGuard}: How Intrinsic Safety is Lost and Found in Diffusion Large Language Models},
  booktitle = {International Conference on Learning Representations (ICLR)},
  year      = {2026},
  doi       = {10.48550/arXiv.2509.24296}
}

@inproceedings{yamabe2026toward,
  author    = {Yamabe, Shojiro and Sakuma, Jun},
  title     = {Toward Safer Diffusion Language Models: Discovery and Mitigation of Priming Vulnerability},
  booktitle = {International Conference on Learning Representations (ICLR)},
  year      = {2026},
  doi       = {10.48550/arXiv.2510.00565}
}

@inproceedings{li2023text,
  author    = {Li, Linyang and Song, Demin and Qiu, Xipeng},
  title     = {Text Adversarial Purification as Defense against Adversarial Attacks},
  booktitle = {Proceedings of the 61st Annual Meeting of the Association for Computational Linguistics (ACL)},
  year      = {2023},
  pages     = {338--350},
  doi       = {10.18653/v1/2023.acl-long.20}
}

@inproceedings{li2025diffusedef,
  author    = {Li, Zhenhao and Zhou, Huichi and Rei, Marek and Specia, Lucia},
  title     = {{DiffuseDef}: Improved Robustness to Adversarial Attacks via Iterative Denoising},
  booktitle = {Proceedings of the 63rd Annual Meeting of the Association for Computational Linguistics (ACL)},
  year      = {2025},
  doi       = {10.48550/arXiv.2407.00248}
}

@inproceedings{gietz2024maskpure,
  author    = {Gietz, Harrison and Kalita, Jugal},
  title     = {{MaskPure}: Improving Defense Against Text Adversaries with Stochastic Purification},
  booktitle = {Natural Language Processing and Information Systems (NLDB)},
  year      = {2024},
  doi       = {10.1007/978-3-031-70239-6_26}
}

@article{zhu2025codefend,
  author  = {Zhu, Fengling and Liu, Boshi and Hua, Jingyu and Zhong, Sheng},
  title   = {{CoDefend}: Cross-Modal Collaborative Defense via Diffusion Purification and Prompt Optimization},
  journal = {arXiv preprint arXiv:2510.11096},
  year    = {2025},
  doi     = {10.48550/arXiv.2510.11096}
}

@article{chen2024diffattack,
  author  = {Chen, Jianqi and Chen, Hao and Chen, Keyan and Zhang, Yilan and Zou, Zhengxia and Shi, Zhenwei},
  title   = {Diffusion Models for Imperceptible and Transferable Adversarial Attack},
  journal = {IEEE Transactions on Pattern Analysis and Machine Intelligence},
  year    = {2024},
  doi     = {10.1109/TPAMI.2024.3372023}
}

@inproceedings{chen2023content,
  author    = {Chen, Zhaoyu and Li, Bo and Wu, Shuang and Ding, Shouhong and Zhang, Wenqiang},
  title     = {Content-based Unrestricted Adversarial Attack},
  booktitle = {Advances in Neural Information Processing Systems},
  year      = {2023},
  doi       = {10.48550/arXiv.2305.10665}
}

@inproceedings{liu2024advdiffusion,
  author    = {Liu, Decheng and Wang, Xijun and Peng, Chunlei and Wang, Nannan and Hu, Ruimin and Gao, Xinbo},
  title     = {Adv-Diffusion: Imperceptible Adversarial Face Identity Attack via Latent Diffusion Model},
  booktitle = {Proceedings of the AAAI Conference on Artificial Intelligence},
  year      = {2024},
  doi       = {10.1609/aaai.v38i4.28067}
}

@article{liu2023diffprotect,
  author  = {Liu, Jiang and Lau, Chun Pong and Chellappa, Rama},
  title   = {{DiffProtect}: Generate Adversarial Examples with Diffusion Models for Facial Privacy Protection},
  journal = {arXiv preprint arXiv:2305.13625},
  year    = {2023},
  doi     = {10.48550/arXiv.2305.13625}
}

@inproceedings{sun2024diffam,
  author    = {Sun, Yuhao and Yu, Lingyun and Xie, Hongtao and Li, Jiaming and Zhang, Yongdong},
  title     = {{DiffAM}: Diffusion-based Adversarial Makeup Transfer for Facial Privacy Protection},
  booktitle = {Proceedings of the IEEE/CVF Conference on Computer Vision and Pattern Recognition (CVPR)},
  year      = {2024},
  doi       = {10.48550/arXiv.2405.09882}
}

@inproceedings{dai2024advdiff,
  author    = {Dai, Xuelong and Liang, Kaisheng and Xiao, Bin},
  title     = {{AdvDiff}: Generating Unrestricted Adversarial Examples using Diffusion Models},
  booktitle = {European Conference on Computer Vision (ECCV)},
  year      = {2024},
  doi       = {10.48550/arXiv.2307.12499}
}

@inproceedings{chen2023advdiffuser,
  author    = {Chen, Xinquan and Gao, Xitong and Zhao, Juanjuan and Ye, Kejiang and Xu, Cheng-Zhong},
  title     = {{AdvDiffuser}: Natural Adversarial Example Synthesis with Diffusion Models},
  booktitle = {Proceedings of the IEEE/CVF International Conference on Computer Vision (ICCV)},
  year      = {2023}
}

@inproceedings{xue2023diffpgd,
  author    = {Xue, Haotian and Araujo, Alexandre and Hu, Bin and Chen, Yongxin},
  title     = {{Diff-PGD}: Diffusion-Based Adversarial Sample Generation for Improved Stealthiness and Controllability},
  booktitle = {Advances in Neural Information Processing Systems},
  year      = {2023},
  doi       = {10.48550/arXiv.2305.16494}
}

@article{huang2025scoreadv,
  author  = {Huang, Chihan and Tang, Hao},
  title   = {{ScoreAdv}: Score-based Targeted Generation of Natural Adversarial Examples via Diffusion Models},
  journal = {arXiv preprint arXiv:2507.06078},
  year    = {2025},
  doi     = {10.48550/arXiv.2507.06078}
}

@article{dai2025semdiff,
  author  = {Dai, Zeyu and Liu, Shengcai and He, Rui and Wu, Jiahao and Lu, Ning and Fan, Wenqi and Li, Qing and Tang, Ke},
  title   = {{SemDiff}: Generating Natural Unrestricted Adversarial Examples via Semantic Attributes Optimization in Diffusion Models},
  journal = {arXiv preprint arXiv:2504.11923},
  year    = {2025},
  doi     = {10.48550/arXiv.2504.11923}
}

@article{collins2025natadiff,
  author  = {Collins, Max and Vice, Jordan and French, Tim and Mian, Ajmal},
  title   = {{NatADiff}: Adversarial Boundary Guidance for Natural Adversarial Diffusion},
  journal = {arXiv preprint arXiv:2505.20934},
  year    = {2025},
  doi     = {10.48550/arXiv.2505.20934}
}

@article{jiang2025apa,
  author  = {Jiang, Kaixun and Chen, Zhaoyu and Guo, Haijing and Li, Jinglun and Fu, Jiyuan and Guo, Pinxue and Tang, Hao and Li, Bo and Zhang, Wenqiang},
  title   = {Enhancing Diffusion-based Unrestricted Adversarial Attacks via Adversary Preferences Alignment},
  journal = {arXiv preprint arXiv:2506.01511},
  year    = {2025},
  doi     = {10.48550/arXiv.2506.01511}
}

@article{chen2023naturalpatch,
  author  = {Chen, Xianyi and Liu, Fazhan and Jiang, Dong and Yan, Kai},
  title   = {Natural Adversarial Patch Generation Method Based on Latent Diffusion Model},
  journal = {arXiv preprint arXiv:2312.16401},
  year    = {2023},
  doi     = {10.48550/arXiv.2312.16401}
}

@article{miao2024advlogo,
  author  = {Miao, Boming and Li, Chunxiao and Zhu, Yao and Sun, Weixiang and Wang, Zizhe and Wang, Xiaoyi and Xie, Chuanlong},
  title   = {{AdvLogo}: Adversarial Patch Attack against Object Detectors based on Diffusion Models},
  journal = {arXiv preprint arXiv:2409.07002},
  year    = {2024},
  doi     = {10.48550/arXiv.2409.07002}
}

@article{wang2024diffpatch,
  author  = {Wang, Zhixiang and Ma, Xingjun and Jiang, Yu-Gang},
  title   = {{BadPatch}: Diffusion-Based Generation of Physical Adversarial Patches},
  journal = {arXiv preprint arXiv:2412.01440},
  year    = {2024},
  doi     = {10.48550/arXiv.2412.01440}
}

@article{wang2025lsdm,
  author  = {Wang, Wenxuan and Qi, Huihui and Huang, Zhixiang and Yin, Bangjie and others},
  title   = {Latent-space diffusion models for stealthy and transferable adversarial attacks on object detection},
  journal = {Neurocomputing},
  volume  = {656},
  pages   = {131456},
  year    = {2025},
  doi     = {10.1016/j.neucom.2025.131456}
}

@article{li2025diffattackx,
  author  = {Li, L. and Zhang, X. and Wang, J. and others},
  title   = {{DiffAttack-X}: An effective transferable adversarial attack based on diffusion models},
  journal = {Applied Intelligence},
  volume  = {55},
  pages   = {1062},
  year    = {2025},
  doi     = {10.1007/s10489-025-06957-6}
}

@inproceedings{kang2024diffattack,
  author    = {Kang, Mintong and Song, Dawn and Li, Bo},
  title     = {{DiffAttack}: Evasion Attacks Against Diffusion-Based Adversarial Purification},
  booktitle = {Advances in Neural Information Processing Systems},
  year      = {2024},
  doi       = {10.48550/arXiv.2311.16124}
}

@article{guo2024advdiffvlm,
  author  = {Guo, Qi and Pang, Shanmin and Jia, Xiaojun and Liu, Yang and Guo, Qing},
  title   = {Efficient Generation of Targeted and Transferable Adversarial Examples for Vision-Language Models Via Diffusion Models},
  journal = {IEEE Transactions on Information Forensics and Security},
  year    = {2024},
  doi     = {10.1109/TIFS.2024.3518072}
}

@inproceedings{xu2024transferable,
  author    = {Xu, Wenzhuo and Chen, Kai and Gao, Ziyi and Wei, Zhipeng and Chen, Jingjing and Jiang, Yu-Gang},
  title     = {Highly Transferable Diffusion-based Unrestricted Adversarial Attack on Pre-trained Vision-Language Models},
  booktitle = {Proceedings of the 32nd ACM International Conference on Multimedia (MM)},
  year      = {2024},
  doi       = {10.1145/3664647.3681538}
}

@inproceedings{zhang2025anyattack,
  author    = {Zhang, Jiaming and Ye, Junhong and Ma, Xingjun and Li, Yige and Yang, Yunfan and Chen, Yunhao and Sang, Jitao and Yeung, Dit-Yan},
  title     = {{AnyAttack}: Towards Large-scale Self-supervised Adversarial Attacks on Vision-Language Models},
  booktitle = {Proceedings of the IEEE/CVF Conference on Computer Vision and Pattern Recognition (CVPR)},
  year      = {2025},
  doi       = {10.48550/arXiv.2410.05346}
}

@inproceedings{li2024hades,
  author    = {Li, Yifan and Guo, Hangyu and Zhou, Kun and Zhao, Wayne Xin and Wen, Ji-Rong},
  title     = {Images Are Achilles' Heel of Alignment: Exploiting Visual Vulnerabilities for Jailbreaking Multimodal Large Language Models},
  booktitle = {European Conference on Computer Vision (ECCV)},
  year      = {2024},
  doi       = {10.48550/arXiv.2403.09792}
}

@inproceedings{liu2024mmsafetybench,
  author    = {Liu, Xin and Zhu, Yichen and Gu, Jindong and Lan, Yunshi and Yang, Chao and Qiao, Yu},
  title     = {{MM-SafetyBench}: A Benchmark for Safety Evaluation of Multimodal Large Language Models},
  booktitle = {European Conference on Computer Vision (ECCV)},
  year      = {2024},
  doi       = {10.48550/arXiv.2311.17600}
}

@inproceedings{wang2025ideator,
  author    = {Wang, Ruofan and Li, Juncheng and Wang, Yixu and Wang, Bo and Wang, Xiaosen and Teng, Yan and Wang, Yingchun and Ma, Xingjun and Jiang, Yu-Gang},
  title     = {{IDEATOR}: Jailbreaking and Benchmarking Large Vision-Language Models Using Themselves},
  booktitle = {Proceedings of the IEEE/CVF International Conference on Computer Vision (ICCV)},
  year      = {2025},
  doi       = {10.48550/arXiv.2411.00827}
}

@inproceedings{zeng2025advi2i,
  author    = {Zeng, Yaopei and Cao, Yuanpu and Cao, Bochuan and Chang, Yurui and Chen, Jinghui and Lin, Lu},
  title     = {{AdvI2I}: Adversarial Image Attack on Image-to-Image Diffusion models},
  booktitle = {International Conference on Machine Learning (ICML)},
  year      = {2025},
  doi       = {10.48550/arXiv.2410.21471}
}

@inproceedings{zhao2023attackvlm,
  author    = {Zhao, Yunqing and Pang, Tianyu and Du, Chao and Yang, Xiao and Li, Chongxuan and Cheung, Ngai-Man and Lin, Min},
  title     = {On Evaluating Adversarial Robustness of Large Vision-Language Models},
  booktitle = {Advances in Neural Information Processing Systems},
  year      = {2023},
  doi       = {10.48550/arXiv.2305.16934}
}

@article{wang2023instructta,
  author  = {Wang, Xunguang and Ji, Zhenlan and Ma, Pingchuan and Li, Zongjie and Wang, Shuai},
  title   = {{InstructTA}: Instruction-Tuned Targeted Attack for Large Vision-Language Models},
  journal = {arXiv preprint arXiv:2312.01886},
  year    = {2023},
  doi     = {10.48550/arXiv.2312.01886}
}

@article{ma2024visual,
  author  = {Ma, Siyuan and Luo, Weidi and Wang, Yu and Liu, Xiaogeng and Chen, Muhao and Li, Bo and Xiao, Chaowei},
  title   = {Visual-RolePlay: Universal Jailbreak Attack on MultiModal Large Language Models via Role-playing Image Character},
  journal = {arXiv preprint arXiv:2405.20773},
  year    = {2024},
  doi     = {10.48550/arXiv.2405.20773}
}

@article{zou2023universal,
  author  = {Zou, Andy and Wang, Zifan and Carlini, Nicholas and Nasr, Milad and Kolter, J.~Zico and Fredrikson, Matt},
  title   = {Universal and Transferable Adversarial Attacks on Aligned Language Models},
  journal = {arXiv preprint arXiv:2307.15043},
  year    = {2023},
  doi     = {10.48550/arXiv.2307.15043}
}

@inproceedings{liu2024autodan,
  author    = {Liu, Xiaogeng and Xu, Nan and Chen, Muhao and Xiao, Chaowei},
  title     = {{AutoDAN}: Generating Stealthy Jailbreak Prompts on Aligned Large Language Models},
  booktitle = {International Conference on Learning Representations (ICLR)},
  year      = {2024},
  doi       = {10.48550/arXiv.2310.04451}
}

@article{chao2023jailbreaking,
  author  = {Chao, Patrick and Robey, Alexander and Dobriban, Edgar and Hassani, Hamed and Pappas, George~J. and Wong, Eric},
  title   = {Jailbreaking Black Box Large Language Models in Twenty Queries},
  journal = {arXiv preprint arXiv:2310.08419},
  year    = {2023},
  doi     = {10.48550/arXiv.2310.08419}
}

@article{paulus2024advprompter,
  author  = {Paulus, Anselm and Zharmagambetov, Arman and Guo, Chuan and Amos, Brandon and Tian, Yuandong},
  title   = {{AdvPrompter}: Fast Adaptive Adversarial Prompting for {LLMs}},
  journal = {arXiv preprint arXiv:2404.16873},
  year    = {2024},
  doi     = {10.48550/arXiv.2404.16873}
}

@inproceedings{qi2024visual,
  author    = {Qi, Xiangyu and Huang, Kaixuan and Panda, Ashwinee and Henderson, Peter and Wang, Mengdi and Mittal, Prateek},
  title     = {Visual Adversarial Examples Jailbreak Aligned Large Language Models},
  booktitle = {Proceedings of the AAAI Conference on Artificial Intelligence},
  year      = {2024},
  doi       = {10.48550/arXiv.2306.13213}
}

@inproceedings{shayegani2024jailbreak,
  author    = {Shayegani, Erfan and Dong, Yue and Abu-Ghazaleh, Nael},
  title     = {Jailbreak in Pieces: Compositional Adversarial Attacks on Multi-Modal Language Models},
  booktitle = {International Conference on Learning Representations (ICLR)},
  year      = {2024},
  doi       = {10.48550/arXiv.2307.14539}
}

@inproceedings{gong2025figstep,
  author    = {Gong, Yichen and Ran, Delong and Liu, Jinyuan and Wang, Conglei and Cong, Tianshuo and Wang, Anyu and Duan, Sisi and Wang, Xiaoyun},
  title     = {{FigStep}: Jailbreaking Large Vision-Language Models via Typographic Visual Prompts},
  booktitle = {Proceedings of the AAAI Conference on Artificial Intelligence},
  year      = {2025},
  doi       = {10.48550/arXiv.2311.05608}
}

@inproceedings{ying2024bap,
  author    = {Ying, Zonghao and Liu, Aishan and Zhang, Tianyuan and Yu, Zhengmin and Liang, Siyuan and Liu, Xianglong and Tao, Dacheng},
  title     = {Jailbreak Vision Language Models via Bi-Modal Adversarial Prompt},
  booktitle = {Advances in Neural Information Processing Systems},
  year      = {2024},
  doi       = {10.48550/arXiv.2406.04031}
}

@inproceedings{wang2024umk,
  author    = {Wang, Ruofan and Ma, Xingjun and Zhou, Hanxu and Ji, Chuanjun and Ye, Guangnan and Jiang, Yu-Gang},
  title     = {White-box Multimodal Jailbreaks against Large Vision-Language Models},
  booktitle = {Proceedings of the 32nd ACM International Conference on Multimedia (MM)},
  year      = {2024},
  doi       = {10.48550/arXiv.2405.17894}
}

@article{rando2024gradient,
  author  = {Rando, Javier and Korevaar, Hannah and Brinkman, Erik and Evtimov, Ivan and Tram{\`e}r, Florian},
  title   = {Gradient-based Jailbreak Images for Multimodal Fusion Models},
  journal = {arXiv preprint arXiv:2410.03489},
  year    = {2024},
  doi     = {10.48550/arXiv.2410.03489}
}

@inproceedings{ho2020denoising,
  author    = {Ho, Jonathan and Jain, Ajay and Abbeel, Pieter},
  title     = {Denoising Diffusion Probabilistic Models},
  booktitle = {Advances in Neural Information Processing Systems},
  year      = {2020},
  doi       = {10.48550/arXiv.2006.11239}
}

@inproceedings{song2021denoising,
  author    = {Song, Jiaming and Meng, Chenlin and Ermon, Stefano},
  title     = {Denoising Diffusion Implicit Models},
  booktitle = {International Conference on Learning Representations (ICLR)},
  year      = {2021},
  doi       = {10.48550/arXiv.2010.02502}
}

@inproceedings{song2021scorebased,
  author    = {Song, Yang and Sohl-Dickstein, Jascha and Kingma, Diederik~P. and Kumar, Abhishek and Ermon, Stefano and Poole, Ben},
  title     = {Score-Based Generative Modeling through Stochastic Differential Equations},
  booktitle = {International Conference on Learning Representations (ICLR)},
  year      = {2021},
  doi       = {10.48550/arXiv.2011.13456}
}

@inproceedings{rombach2022high,
  author    = {Rombach, Robin and Blattmann, Andreas and Lorenz, Dominik and Esser, Patrick and Ommer, Bj{\"o}rn},
  title     = {High-Resolution Image Synthesis with Latent Diffusion Models},
  booktitle = {Proceedings of the IEEE/CVF Conference on Computer Vision and Pattern Recognition (CVPR)},
  year      = {2022},
  doi       = {10.48550/arXiv.2112.10752}
}

@inproceedings{austin2021structured,
  author    = {Austin, Jacob and Johnson, Daniel and Ho, Jonathan and Tarlow, Daniel and van~den Berg, Rianne},
  title     = {Structured Denoising Diffusion Models in Discrete State-Spaces},
  booktitle = {Advances in Neural Information Processing Systems},
  year      = {2021},
  doi       = {10.48550/arXiv.2107.03006}
}

@inproceedings{lou2024discrete,
  author    = {Lou, Aaron and Meng, Chenlin and Ermon, Stefano},
  title     = {Discrete Diffusion Modeling by Estimating the Ratios of the Data Distribution ({SEDD})},
  booktitle = {International Conference on Machine Learning (ICML)},
  year      = {2024},
  doi       = {10.48550/arXiv.2310.16834}
}

@article{shi2024simplified,
  author  = {Shi, Jiaxin and Han, Kehang and Wang, Zhe and Doucet, Arnaud and Titsias, Michalis},
  title   = {Simplified and Generalized Masked Diffusion for Discrete Data ({MDLM})},
  journal = {arXiv preprint arXiv:2406.04329},
  year    = {2024},
  doi     = {10.48550/arXiv.2406.04329}
}

@article{nie2025llada,
  author  = {Nie, Shen and Zhu, Fengqi and You, Zebin and Zhang, Xiaolu and Ou, Jingyang and Hu, Jun and Zhou, Jun and Lin, Yankai and Wen, Ji-Rong and Li, Chongxuan},
  title   = {Large Language Diffusion Models ({LLaDA})},
  journal = {arXiv preprint arXiv:2502.09992},
  year    = {2025},
  doi     = {10.48550/arXiv.2502.09992}
}

@inproceedings{nichol2022glide,
  author    = {Nichol, Alexander~Q. and Dhariwal, Prafulla and Ramesh, Aditya and Shyam, Pranav and Mishkin, Pamela and McGrew, Bob and Sutskever, Ilya and Chen, Mark},
  title     = {{GLIDE}: Towards Photorealistic Image Generation and Editing with Text-Guided Diffusion Models},
  booktitle = {International Conference on Machine Learning (ICML)},
  year      = {2022},
  doi       = {10.48550/arXiv.2112.10741}
}

@inproceedings{dhariwal2021diffusion,
  author    = {Dhariwal, Prafulla and Nichol, Alexander~Q.},
  title     = {Diffusion Models Beat {GAN}s on Image Synthesis},
  booktitle = {Advances in Neural Information Processing Systems},
  year      = {2021},
  doi       = {10.48550/arXiv.2105.05233}
}

@inproceedings{achiam2023gpt4,
  author  = {Achiam, Josh and Adler, Steven and Agarwal, Sandhini and Ahmad, Lama and Akkaya, Ilge and Aleman, Florencia~Leoni and Almeida, Diogo and Altenschmidt, Janko and Altman, Sam and Anadkat, Shyamal and others},
  title   = {{GPT-4} Technical Report},
  journal = {arXiv preprint arXiv:2303.08774},
  year    = {2023},
  doi     = {10.48550/arXiv.2303.08774}
}

@article{touvron2023llama2,
  author  = {Touvron, Hugo and Martin, Louis and Stone, Kevin and Albert, Peter and Almahairi, Amjad and Babaei, Yasmine and Bashlykov, Nikolay and Batra, Soumya and Bhargava, Prajjwal and Bhosale, Shruti and others},
  title   = {{LLaMA} 2: Open Foundation and Fine-Tuned Chat Models},
  journal = {arXiv preprint arXiv:2307.09288},
  year    = {2023},
  doi     = {10.48550/arXiv.2307.09288}
}

@article{jabbar2025red,
  author  = {Jabbar, Muhammad~Shahid and Al-Azani, Sadam and Alotaibi, Abrar and Ahmed, Moataz},
  title   = {Red Teaming Large Language Models: A Comprehensive Review and Critical Analysis},
  journal = {Preprint submitted to Elsevier},
  year    = {2025}
}

@article{lin2025red,
  author  = {Lin, Lizhi and Mu, Honglin and Zhai, Zenan and Wang, Minghan and Wang, Yuxia and Wang, Renxi and Gao, Junjie and Zhang, Yixuan and Che, Wanxiang and Baldwin, Timothy and Han, Xudong and Li, Haonan},
  title   = {Against The Achilles' Heel: A Survey on Red Teaming for Generative Models},
  journal = {Journal of Artificial Intelligence Research},
  year    = {2025},
  doi     = {10.48550/arXiv.2404.00629}
}

@article{cui2024risk,
  author  = {Cui, Tianyu and Wang, Yanling and Fu, Chuanpu and Xiao, Yong and Li, Sijia and Deng, Xinhao and Liu, Yunpeng and Zhang, Qinglin and Qiu, Ziyi and Li, Peiyang and Tan, Zhixing and Xiong, Junwu and Kong, Xinyu and Wen, Zujie and Xu, Ke and Li, Qi},
  title   = {Risk Taxonomy, Mitigation, and Assessment Benchmarks of Large Language Model Systems},
  journal = {arXiv preprint arXiv:2401.05778},
  year    = {2024},
  doi     = {10.48550/arXiv.2401.05778}
}

@article{liu2024jailbreaking,
  author  = {Liu, Yi and Deng, Gelei and Xu, Zhengzi and Li, Yuekang and Zheng, Yaowen and Zhang, Ying and Zhao, Lida and Zhang, Tianwei and Wang, Kailong and Liu, Yang},
  title   = {Jailbreaking ChatGPT via Prompt Engineering: An Empirical Study},
  journal = {arXiv preprint arXiv:2305.13860},
  year    = {2024},
  doi     = {10.48550/arXiv.2305.13860}
}

@article{yi2024jailbreak,
  author  = {Yi, Sibo and Liu, Yule and Sun, Zhen and Cong, Tianshuo and He, Xinlei and Song, Jiaxing and Xu, Ke and Li, Qi},
  title   = {Jailbreak Attacks and Defenses Against Large Language Models: A Survey},
  journal = {arXiv preprint arXiv:2407.04295},
  year    = {2024},
  doi     = {10.48550/arXiv.2407.04295}
}

@article{huang2024survey,
  author  = {Huang, Xiaowei and Ruan, Wenjie and Huang, Wei and Jin, Gaojie and Dong, Yi and Wu, Changshun and Bensalem, Saddek and Mu, Ronghui and Qi, Yi and Zhao, Xingyu and Cai, Kaiwen and Zhang, Yanghao and Wu, Sihao and Xu, Peipei and Wu, Dengyu and Freitas, Andre and Mustafa, Mustafa~A.},
  title   = {A Survey of Safety and Trustworthiness of Large Language Models through the Lens of Verification and Validation},
  journal = {Artificial Intelligence Review},
  year    = {2024},
  doi     = {10.1007/s10462-024-10884-2}
}

@inproceedings{mazeika2024harmbench,
  author    = {Mazeika, Mantas and Phan, Long and Yin, Xuwang and Zou, Andy and Wang, Zifan and Mu, Norman and Sakhaee, Elham and Li, Nathaniel and Basart, Steven and Li, Bo and Forsyth, David and Hendrycks, Dan},
  title     = {{HarmBench}: A Standardized Evaluation Framework for Automated Red Teaming and Robust Refusal},
  booktitle = {International Conference on Machine Learning (ICML)},
  year      = {2024},
  doi       = {10.48550/arXiv.2402.04249}
}

@article{chao2024jailbreakbench,
  author  = {Chao, Patrick and Debenedetti, Edoardo and Robey, Alexander and Andriushchenko, Maksym and Croce, Francesco and Sehwag, Vikash and Dobriban, Edgar and Flammarion, Nicolas and Pappas, George~J. and Tram{\`e}r, Florian and Hassani, Hamed and Wong, Eric},
  title   = {{JailbreakBench}: An Open Robustness Benchmark for Jailbreaking Large Language Models},
  journal = {arXiv preprint arXiv:2404.01318},
  year    = {2024},
  doi     = {10.48550/arXiv.2404.01318}
}

@inproceedings{gehman2020realtoxicity,
  author    = {Gehman, Samuel and Gururangan, Suchin and Sap, Maarten and Choi, Yejin and Smith, Noah~A.},
  title     = {{RealToxicityPrompts}: Evaluating Neural Toxic Degeneration in Language Models},
  booktitle = {Findings of the Association for Computational Linguistics: EMNLP},
  year      = {2020},
  doi       = {10.18653/v1/2020.findings-emnlp.301}
}

@inproceedings{deng2009imagenet,
  author    = {Deng, Jia and Dong, Wei and Socher, Richard and Li, Li-Jia and Li, Kai and Fei-Fei, Li},
  title     = {{ImageNet}: A Large-Scale Hierarchical Image Database},
  booktitle = {Proceedings of the IEEE/CVF Conference on Computer Vision and Pattern Recognition (CVPR)},
  year      = {2009}
}

@inproceedings{lin2014mscoco,
  author    = {Lin, Tsung-Yi and Maire, Michael and Belongie, Serge and Hays, James and Perona, Pietro and Ramanan, Deva and Doll{\'a}r, Piotr and Zitnick, C.~Lawrence},
  title     = {Microsoft {COCO}: Common Objects in Context},
  booktitle = {European Conference on Computer Vision (ECCV)},
  year      = {2014}
}

@article{perez2022red,
  author  = {Perez, Ethan and Huang, Saffron and Song, Francis and Cai, Trevor and Ring, Roman and Aslanides, John and Glaese, Amelia and McAleese, Nat and Irving, Geoffrey},
  title   = {Red Teaming Language Models with Language Models},
  journal = {arXiv preprint arXiv:2202.03286},
  year    = {2022},
  doi     = {10.48550/arXiv.2202.03286}
}

@inproceedings{ganguli2022red,
  author    = {Ganguli, Deep and Lovitt, Liane and Kernion, Jackson and Askell, Amanda and Bai, Yuntao and Kadavath, Saurav and Mann, Ben and Perez, Ethan and Schiefer, Nicholas and Ndousse, Kamal and others},
  title     = {Red Teaming Language Models to Reduce Harms: Methods, Scaling Behaviors, and Lessons Learned},
  booktitle = {arXiv preprint arXiv:2209.07858},
  year      = {2022},
  doi       = {10.48550/arXiv.2209.07858}
}

@inproceedings{wei2023jailbroken,
  author    = {Wei, Alexander and Haghtalab, Nika and Steinhardt, Jacob},
  title     = {Jailbroken: How Does {LLM} Safety Training Fail?},
  booktitle = {Advances in Neural Information Processing Systems},
  year      = {2023},
  doi       = {10.48550/arXiv.2307.02483}
}

@inproceedings{robey2023smoothllm,
  author    = {Robey, Alexander and Wong, Eric and Hassani, Hamed and Pappas, George~J.},
  title     = {{SmoothLLM}: Defending Large Language Models against Jailbreaking Attacks},
  booktitle = {arXiv preprint arXiv:2310.03684},
  year      = {2023},
  doi       = {10.48550/arXiv.2310.03684}
}

@article{inan2023llamaguard,
  author  = {Inan, Hakan and Upasani, Kartikeya and Chi, Jianfeng and Rungta, Rashi and Iyer, Krithika and Mao, Yuning and Tontchev, Michael and Hu, Qing and Fuller, Brian and Testuggine, Davide and Khabsa, Madian},
  title   = {Llama Guard: {LLM}-based Input-Output Safeguard for Human-{AI} Conversations},
  journal = {arXiv preprint arXiv:2312.06674},
  year    = {2023},
  doi     = {10.48550/arXiv.2312.06674}
}

@article{page2021prisma,
  author  = {Page, Matthew~J. and McKenzie, Joanne~E. and Bossuyt, Patrick~M. and Boutron, Isabelle and Hoffmann, Tammy~C. and Mulrow, Cynthia~D. and Shamseer, Larissa and Tetzlaff, Jennifer~M. and Akl, Elie~A. and Brennan, Sue~E. and others},
  title   = {The {PRISMA} 2020 statement: an updated guideline for reporting systematic reviews},
  journal = {BMJ},
  year    = {2021},
  volume  = {372},
  doi     = {10.1136/bmj.n71}
}

@article{jin2020bertattack,
  author  = {Jin, Di and Jin, Zhijing and Zhou, Joey~Tianyi and Szolovits, Peter},
  title   = {Is {BERT} Really Robust? A Strong Baseline for Natural Language Attack on Text Classification and Entailment ({TextFooler})},
  journal = {Proceedings of the AAAI Conference on Artificial Intelligence},
  year    = {2020},
  doi     = {10.1609/aaai.v34i05.6311}
}

@article{mehrabi2023flirt,
  title   = {{FLIRT}: Feedback Loop In-context Red Teaming},
  author  = {Mehrabi, Ninareh and Goyal, Palash and Dupuy, Christophe and Hu, Qian and Ghosh, Shalini and Zemel, Richard and Chang, Kai-Wei and Galstyan, Aram and Gupta, Rahul},
  journal = {arXiv preprint arXiv:2308.04265},
  year    = {2023}
}

@article{landis1977measurement,
  title   = {The measurement of observer agreement for categorical data},
  author  = {Landis, J. Richard and Koch, Gary G.},
  journal = {Biometrics},
  volume  = {33},
  number  = {1},
  pages   = {159--174},
  year    = {1977},
  publisher = {JSTOR}
}

@inproceedings{nie2022diffusion,
  title     = {Diffusion Models for Adversarial Purification},
  author    = {Nie, Weili and Guo, Brandon and Huang, Yujia and Xiao, Chaowei and Vahdat, Arash and Anandkumar, Anima},
  booktitle = {Proceedings of the 39th International Conference on Machine Learning (ICML)},
  year      = {2022}
}

\end{document}